\documentclass[twocolumn,english,showpacs,amsmath]{revtex4-1}
\usepackage{graphicx}
\usepackage{amssymb}

\newcommand{\figref}[1]{Fig.\,\ref{#1}}
\newcommand{\eqrefE}[1]{Eqn.\,\eqref{#1}}

\begin{document}

\title{Kinetics of a single trapped ion in an ultracold buffer gas }

\author{Christoph Zipkes}
\email{cdz22@cam.ac.uk}
\author{Lothar Ratschbacher}
\author{Carlo Sias}
\author{Michael K{\"o}hl}
\affiliation{Cavendish Laboratory, University of Cambridge, JJ Thomson Avenue, Cambridge CB3 0HE, United Kingdom}

\begin{abstract}
The immersion of a single ion confined by a radiofrequency trap in an ultracold atomic gas extends the concept of buffer gas cooling to a new temperature regime. The steady state energy distribution of the ion is determined by its kinetics in the radiofrequency field rather than the temperature of the buffer gas. Moreover, the finite size of the ultracold gas facilitates the observation of back-action of the ion onto the buffer gas. We numerically investigate the system's properties depending on atom-ion mass ratio, trap geometry, differential cross-section, and non-uniform neutral atom density distribution. 
Experimental results are well reproduced by our model considering only elastic collisions. 
We identify excess micromotion to set the typical scale for the ion energy statistics and explore the applicability of the mobility collision cross-section to the ultracold regime.
\end{abstract}


\date{\today}

\maketitle

\section{Introduction}
\label{sec:introduction}
Trapped ion systems are among the most promising candidates for quantum information processing~\cite{Cirac1995,Blatt2008}, precision measurements~\cite{Blatt1992}, and quantum chemistry~\cite{Krems2008}. For many of these applications it is required to cool the ions to low temperatures.
To this end, various techniques such as laser cooling, resistive cooling, sympathetic cooling by other ions, or buffer gas cooling are routinely used. 
Ultracold atomic gases have recently become available in hybrid systems with trapped ions~\cite{Cetina2007,Grier2009,Zipkes2010,Zipkes2010b,Schmid2010,Ravi2010}, extending the concept of buffer gas cooling to ultralow temperatures.
Understanding this new regime and how it relates to conventional buffer gas cooling is essential for any future application to trapped ions.

Cooling of an ion in a buffer gas is caused by elastic collisions.
They are dominated by the long range polarization interaction, and cross-sections are considerably larger than for collisions between two neutral atoms~\cite{Cote2000,Cote2002,Idziaszek2007,Kollath2007,Idziaszek2009,Sherkunov2009,Doerk2010,Gao2010,Goold2010,Idziaszek2010,Krych2010}. Therefore, collision rates are large and cooling is expected to be efficient~\cite{Makarov2003}. Moreover, it has been proposed that internal degrees of freedom of ionic molecules could also be cooled by using an ultracold buffer gas~\cite{Hudson2009}. Another specific feature of the polarization interaction is that collisions affecting the ion's mobility happen with rates independent of the collision energy~\cite{Langevin1905}. This can lead to simplified system behaviour and has, for example, been applied in ion mobility spectrometry~\cite{Albritton1968}.

The motion of an ion in a radiofrequency (RF) trap can be decomposed into a fast driven motion, the micromotion, and a slow secular motion.
In every collision the energy of the RF-field couples via the micromotion to the neutral atom's energy and the ion's secular energy. This can lead to ion energy removal or intake, sensitive to the RF-phase in the moment of the collision~\cite{Major1968,Ridinger2007,Ridinger2009}. The average effect of many collisions results in cooling or heating and depends on parameters such as the ratio between the mass of the ion $m_{i}$ and the mass of the neutral atom $m_{n}$. For very heavy neutrals runaway heating of the ion is expected~\cite{Major1968}, whereas very light neutrals enable efficient buffer gas cooling. In the mass ratio regime between the two extremes increased ion trap loss can be observed~\cite{Moriwaki1998,Green2007}. This effect has been explained by non-thermal energy distributions of the ion obtained from Monte Carlo simulations~\cite{Devoe2009}. Such numerical simulations are a well established tool to model a trapped ion interacting with a buffer gas~\cite{Kim1997,Kellerbauer2001,Schwarz2006}. They can account for energy dependent scattering rates, complex electric field geometries, and other experimental parameters, which are difficult to treat analytically. In previous calculations, buffer gases at ambient temperatures have been assumed, and the cooling of the ion has been limited to the buffer gas temperature. However, in the recent experiments with ultracold neutral buffer gases a new energy scale related to excess micromotion has become dominant. The direct relation between the ion's mean energy and the excess micromotion has been observed in~\cite{Zipkes2010b}, for a system of $\text{Yb}^+ - \text{Rb}$.

Here, we investigate the kinetics of a single ion colliding elastically with an ultracold buffer gas, by applying Monte Carlo techniques. The effects on the ultracold neutral cloud are modelled using a semiclassical differential cross-section. 
The results on neutral atom loss and temperature increase, and the dependence of ion energy on excess micromotion are in good agreement with experiments. 

The paper is organized as follows: In section~\ref{sec:simModel} we describe the basic simulation procedure and the underlying physical model. The classical Langevin interaction model is applied in section~\ref{sec:LangevinScattering} to derive the ion's energy statistics depending on the mass ratio, trap geometry, and scattering rate. In section~\ref{sec:secmiclassicalScattering} we explain effects on the neutral atom cloud as a result of the energy dependent differential cross-section and the non-uniform neutral density distribution.

\section{Simulation model}
\label{sec:simModel}
The time evolution of a trapped single ion colliding with ultracold atoms is modelled using a simulation consisting of an analytical and a numerical part. In the time between collisions, trajectories are analytically described using the pseudo-potential approximation, while elastic collisions are taken into account using Monte Carlo techniques. 
 
\subsection{Ion trajectory}
\label{sec:ionTrajectory}

We consider a single ion confined by the RF-quadrupole potential of a linear Paul trap
\begin{equation}
 \Phi_{RF} = V_0\,\frac{x^2-y^2}{2\,{R_T}^2}\,\sin(\Omega_T\,t)	\;.
 \label{eqn:phiRf}
\end{equation}
Here, $V_0$ is the RF-voltage amplitude applied with frequency $\Omega_T$ to electrodes at a distance ${R_T}$ away from the trap symmetry axis. In addition, we consider a static quadrupole potential confining the ion along the trap symmetry axis with trapping frequency $\omega_z$,
\begin{equation}
 \Phi_{static} = \frac{m_{i}}{2\,Q}\,\omega_z^2\,\left(z^2-\frac{1}{2}(x^2+y^2)\right)	\;. 
 \label{eqn:phiStatic}
\end{equation}
Q is the ion's charge.
Mathieu equations describe the classical motion of an ion in the combined potential (see for example~\cite{Wineland1998,Leibfried2003}), 
using the parameters $a=2\,\frac{\omega_z^2}{\Omega_T^2}$ and $q=\sqrt{8}\,\frac{\omega_p}{\Omega_T}$ with $\omega_p=\frac{Q}{\sqrt{2}\,m_{i}}\,\frac{V_0}{{R_T}^2\,\Omega_T}$.  
For $a < q^2/2 \ll 1$ the Floquet solution to first order in $q$ yields 
\begin{subequations}\label{eqn:rIon}
\begin{align}
 r_{ion,x}&=  A_x\,\sin(\omega_x\,t+\varphi_x) \,\left[1+\frac{q}{2}\sin(\Omega_T\,t)\right]\\
 r_{ion,y}&=  A_y\,\sin(\omega_y\,t+\varphi_y) \,\left[1-\frac{q}{2}\sin(\Omega_T\,t)\right] \;,
\end{align}
\end{subequations}
which is usually referred to as the pseudo-potential approximation. It consists of a rapidly oscillating micromotion term and the secular motion, which is harmonic with frequencies $\omega_{x,y}=\sqrt{\omega_p^2-\frac{1}{2}\omega_z^2}$ and amplitudes $A_{x,y}=\frac{1}{\omega_{x,y}}\sqrt{\frac{2\,E_{x,y}}{m_{i}}}$. A full secular trajectory $\vec{r}_{sec}$ including the harmonic motion along the trap symmetry axis is described by three energies $E_j$ and three phases $\varphi_j$, $j\in\{x,y,z\}$, with
\begin{equation}
 r_{sec,j} = \frac{1}{\omega_j}\sqrt{\frac{2\,E_j}{m_{i}}}\,\sin(\omega_j\,t+\varphi_j)	\;.
 \label{eqn:posIonTa}
\end{equation}
This formula, describing a three-dimensional harmonic oscillator, will be used throughout the following calculations to approximate the ion's position.
The total secular energy $E_x+E_y+E_z$ will be referred to as the ion energy.

The motion of the ion is affected by collisions with the neutral atoms. We assume them to be instantaneous, meaning, the timescale of the collision is shorter than $\Omega_T^{-1}$, which is the shortest timescale of the motion of the ion. This assumption implies that every collision is sensitive to the momentary relative velocity, including the micromotion~\footnote{This assumption is valid for collision energies above $\hbar\,\Omega_T$.}. Therefore, we consider a number of effects causing contributions to the micromotion. Firstly, the intrinsic micromotion described in \eqrefE{eqn:rIon} is proportional to the distance of the ion from the centre of the RF-quadrupole field. Secondly, static offset electric fields displace the minimum of the ion trapping potential by a distance $(\Delta x, \Delta y)$ from the symmetry axis of \eqrefE{eqn:phiRf}. Thirdly, RF pickup on end-cap electrodes can lead to micromotion along the trap symmetry axis. These contributions are summed in the expression
\begin{equation}
	\vec{v}_{mm}= \sqrt{2}\,\omega_p
	\left( \begin{array}{c}	r_{sec,x}(t)+\Delta x  \\	-r_{sec,y}(t)-\Delta y \\	c_z 
	\end{array} \right)
	\cos(\Omega_T\,t)		\;,
	\label{eqn:ionMicromotionVelocity1}
\end{equation}
where $c_z$ parameterizes the micromotion along the trap symmetry axis~\footnote{Not included in this expression are possible additional micromotion terms that are $\pi/2$ out of phase, therefore proportional to $\sin(\Omega_T\,t)$, which can arise, for example, from RF-phase mismatches on opposing electrodes. They are added to $\vec{v}_{mm}$ for simulations where specific experimental data is to be represented~\cite{Zipkes2010b}.}.

The ion velocity considered for collisions is
\begin{equation}
	\vec{v}_{ion} = \vec{v}_{sec} + \vec{v}_{mm}
	 \; ,
	\label{eqn:ionMotionVelocity1}
\end{equation}
with $\vec{v}_{sec}=\frac{\mathrm{d}}{\mathrm{d}t}\vec{r}_{sec}$. This is similar to the time derivative of \eqrefE{eqn:rIon} but includes all the excess micromotion terms.

\subsection{Collision dynamics}
\label{sec:collDynamics}
The simulation uses classical trajectories for the motion of the ion. Therefore, its validity is restricted to ion energies well above the energy quanta of secular motion $\hbar\omega$. 
The temperature of the ultracold buffer gas is assumed to be well below this energy scale, and in the collision the neutral atom's initial energy is neglected. 
Due to conservation of energy and momentum, the elastic scattering process is defined by the scattering angles $(\theta,\phi)$. 
The ion's velocity changes according to
\begin{equation}
	\vec{v}_{ion,f}= (1-\beta)\,\vec{v}_{ion,i}+\beta\,\mathcal{R}\,\vec{v}_{ion,i} 	\;,
	\label{eqn:mcDeflection}
\end{equation}
with $\vec{v}_{ion,i}$ ($\vec{v}_{ion,f}$) being the initial (final) velocity given by \eqrefE{eqn:ionMotionVelocity1} at the time of the collision, $\beta=\frac{m_{n}}{m_{i}+m_{n}}$ and $\mathcal{R}$ is the rotation matrix determined by $\theta$ and $\phi$, with respect to the direction of $\vec{v}_{ion,i}$. From $\vec{v}_{sec,f}$ and $\vec{r}_{sec}$ a new set of $\varphi_j$ and $E_j$ can be determined, which describes the ion's trajectory after the collision. 

To illustrate its impact on the motion of the ion, \eqrefE{eqn:mcDeflection} can be rewritten in terms of the secular velocity, yielding
\begin{equation}
	\vec{v}_{sec,f}= (1-\beta)\,\vec{v}_{sec,i} + \beta\,\mathcal{R}\,\vec{v}_{sec,i} + \beta\,(\mathcal{R}-\textbf{1})\,\vec{v}_{mm} 	\;.
	\label{eqn:mcDeflection4}
\end{equation}
This expression shows how $\vec{v}_{mm}$ couples to the secular velocity in every collision. Note that $\vec{v}_{mm}$ is, by definition of \eqrefE{eqn:ionMicromotionVelocity1}, the same before and after the collision since it only depends on the position $\vec{r}_{sec}$ and time $t$ of the instantaneous collision.

\subsection{Scattering rate}
\label{sec:scattRate}
The probability $\mathrm{d}P_c$ for the ion to collide with a neutral atom within a short time interval $\mathrm{d}t$ defines the scattering rate
\begin{equation}
	\Gamma(t) = \frac{\mathrm{d}P_c}{\mathrm{d}t} = n(\vec{x})\,\sigma(E_c)\,v_{ion}(t) 	\;.
	\label{eqn:scatteringRate}
\end{equation}
It is proportional to the neutral atom density $n(\vec{x})$ at the ion's position. The cross-section $\sigma(E_c)$ is usually a function of the collision energy $E_c$, which, neglecting the energy of the neutral atom, is given by $E_c=\beta\,\frac{m_{i}}{2}\,v_{ion}^2$. 

The ion's position changes on a timescale of $\omega_j^{-1}$ while the velocity of the ion $v_{ion}$ changes on a timescale of $\Omega_T^{-1}$. In general, $\Gamma(t)$ will therefore be time dependent in a non-trivial way. The sampling method used to efficiently choose the time of collision is explained in Appendix~\ref{app:collTimeSampling}.

A technical description of the main simulation loop is given in Appendix~\ref{app:simulationLoop}.

\subsection{Inelastic collisions}
\label{sec:inelasticCollisions}
Inelastic processes like charge exchange, spin exchange or molecule formation have been predicted to occur in the hybrid system~\cite{Massey1934,Cote2000,Cote2000b,Makarov2003,Zhang2009}. 
In experiments~\cite{Grier2009,Zipkes2010,Zipkes2010b,Schmid2010}, charge exchange, which is typically signaled by the loss of the ion, has been observed at rates many orders of magnitude lower than the elastic collision rate, in the non-resonant case. 
Spin exchange collisions can occur with rates comparable to elastic scattering~\cite{Makarov2003}, and energy from internal states can be transferred to the external degrees of freedom. In a spin-stretched configuration, however, spin exchange is suppressed. 

In our simulation, we can include inelastic effects by introducing additional, competing rates, defined as in \eqrefE{eqn:scatteringRate}, but with inelastic cross-sections $\sigma_i(E_c)$ instead. The effect of any inelastic event on the hybrid system depends primarily on the question whether the original ion still exists after the process. If this is not the case, the simulation can be stopped at the first occurrence. If the ion continues to exist, the amount of energy released or absorbed by the internal states of the colliding particles needs to be considered in a modified version of \eqrefE{eqn:mcDeflection}. In either case, our simulation is able to predict the rate at which inelastic events occur, given the inelastic cross-section $\sigma_i(E_c)$. On the other hand, the simulation can be used, if a rate is measured experimentally, to determine $\sigma_i(E_c)$. 

In the following sections we consider elastic processes only, assuming inelastic processes either to happen very rarely, in line with the experimens~\cite{Zipkes2010,Zipkes2010b,Schmid2010}, 
or to involve only small amounts of internal energy, which do not significantly affect the system. 

\section{Langevin scattering}
\label{sec:LangevinScattering}
For the motion of an ion in a neutral gas, mainly large angle scattering is considered relevant, as small deflections do not significantly change the ion's trajectory. This assumption leads to the Langevin scattering model, which successfully describes the ion's mobility in previous experiments with ions in a neutral buffer gas. Here, we apply the Langevin scattering model to the trapped ion system including the effects of micromotion. We investigate the properties of the ion's energy and will later compare these results with those from a more complete semiclassical scattering model, in section~\ref{sec:secmiclassicalScattering}. We will indeed find good agreement between the two models in describing the energy distribution of the ion, confirming the above assumption, that large angle scattering events determine the evolution of the ion's energy.

The ion-neutral interaction is dominated by the attractive polarization interaction, which is of the form
\begin{equation}
  V(R) = -\frac{C_4}{2 R^4}
	\label{eqn:dxs_C4_potential}
\end{equation} 
with $C_4=\alpha_0 Q^2/(4 \pi \epsilon_0)^2$ being proportional to the neutral particle polarizability $\alpha_0$. $R$ is the internuclear separation. 
Classically one can define a critical impact parameter $b_c=(2\,C_4/E_c)^{1/4}$~\cite{Vogt1954}. Collisions with impact parameter $b>b_c$ lead to small deflections and are neglected. Impact parameters $b<b_c$ result in inward-spiralling trajectories, which lead to almost uniformly distributed scattering angles into all directions as in hard-sphere scattering. The resulting cross-section for large angle scattering, $\sigma_L=\pi\,b_c^2$ is proportional to the inverse collision velocity, and leads to a scattering rate independent of the collision energy~\cite{Langevin1905}. 

\subsection{Energy scale}
\label{sec:energyScale}
Our aim is to determine the energy scale of the ion on its classical trajectory after many collisions such that the initial conditions for $E_j$ can be neglected. Neither the Langevin scattering at its energy-independent rate, nor the ultracold neutral buffer gas, which we assume at $T=0$, introduce such a scale.
As a consequence the only energy scale in the system is defined by the excess micromotion, see \eqrefE{eqn:ionMicromotionVelocity1}. 
This is in contrast to the case of a buffer gas with non-negligible temperature, where the energy scale is rather set by the temperature of the neutral gas~\cite{Devoe2009}. 

To associate the excess-micromotion with an energy scale we define
\begin{equation}
 E_{mm,e} = \frac{m_{i}}{2} v_{mm,e}^2
 \label{eqn:hardSpere_Emm}
\end{equation}
with $v_{mm,e}$ being the full velocity amplitude of the micromotion for an ion in the centre of the ion trapping potential. $v_{mm,e}$ depends on the displacement $(\Delta x, \Delta y)$ caused by the uncompensated offset electric field. Note that when the offset electric field is compensated using a photon correlation measurement~\cite{Bluemel1989,Berkeland1998}, photon shot noise usually limits the lowest achievable $E_{mm,e}$. 

Since $E_{mm,e}$ is the only energy scale in the colliding system it is sufficient to express all energies in units of $E_{mm,e}$. This gives general results for any amplitude of uncompensated micromotion.
It also implies that any statistical measure of ion energy has to scale with $E_{mm,e}$ and therefore with the square of the excess micromotion amplitude. Such quadratic dependence of the mean ion energy has been experimentally observed in~\cite{Zipkes2010b}.

\subsection{Energy spectrum}
In order to treat scattering in the presence of micromotion in its most general form we choose all trap frequencies ($\omega_{x,y,z}, \Omega_T$) to have irrational ratios. We assume a very low homogeneous neutral atom density such that the scattering rate $\Gamma$ is much smaller than the trap frequencies. This ensures that two consecutive collisions happen at uncorrelated positions. 

\begin{figure}[b]
    \includegraphics{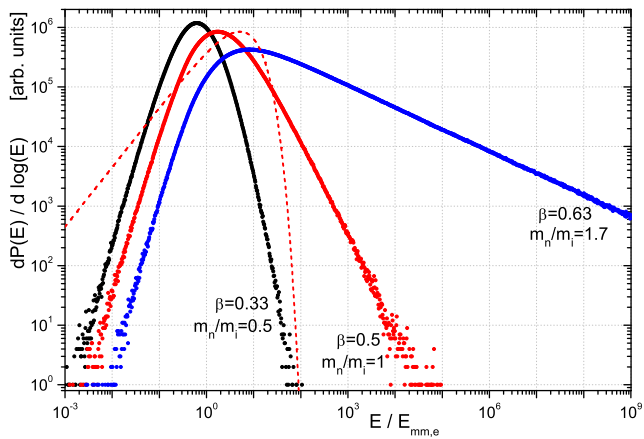}
	\caption[Hard sphere scattering energy spectrum]{\label{fig:langevinThermal} Energy spectra of the ion for three different mass ratios. $10^8$~energies are sampled into logarithmically spaced bins with an energy resolution of 100 bins per decade. We choose $\Delta x=\Delta y$ and $c_z=0$ and trap parameters $\frac{q^2}{a}=50$. In the case where the ion mass is twice the neutral mass (black) the ion's mean energy is $0.8\,E_{mm,e}$, in the case for equal masses (red) $5\,E_{mm,e}$. The red dashed line corresponds to a thermal energy distribution with $5\,E_{mm,e}$. The spectrum for a lighter ion, $\frac{m_{n}}{m_{i}} = 1.7$ (blue), contains a significant contribution of very high energies, typically leading to quick ion loss due to finite trap depth. }
\end{figure}

We obtain the energy spectrum from the simulation by binning the secular energy after each collision on a logarithmic scale. This measures a logarithmic energy probability distribution $\mathrm{d}P(E)/\mathrm{d}\log(E)$. 
\figref{fig:langevinThermal} shows such energy spectra for three different mass ratios of the atom and the ion. 
For heavier neutral atoms the mean energy of the ion increases and a tail in the spectrum towards higher energies becomes dominant. 
Even for equal masses ($\beta = 0.5$) the tail towards high energies is evident when compared to the thermal distribution with the same mean energy. 
For any mass ratio, the obtained spectrum distinctly differs from a thermal distribution, also because very low energies ($E \ll E_{mm,e}$) are extremely rare. This supports the validity of the classical treatment of trajectories and instantaneous collisions ($E>\hbar\Omega_T$). 

\begin{figure}[ht]
    \includegraphics{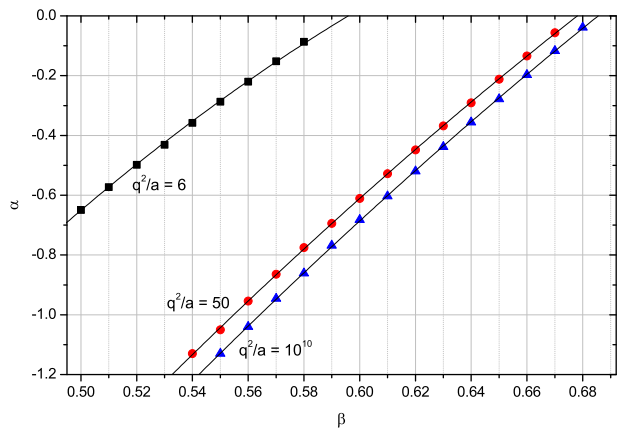}
	\caption[Critical Beta for Different Geometries]{\label{fig:betaCritVsQSqByA}The exponents $\alpha$ obtained from fitting power laws to the high energy tails of the ion energy spectra are plotted against the mass ratio coefficient $\beta$. This is done for three different trap geometries, for a spherical trap ($\frac{q^2}{a}=6$, black rectangles), for an elongated trap ($\frac{q^2}{a}=50$, red circles) and for the extreme case with negligible axial confinement ($\frac{q^2}{a}=10^{10}$, blue triangles). The data is fitted with second order polynomials to extrapolate to $\beta_{crit}$ for which the exponent $\alpha$ becomes 0.  } 
\end{figure}

A power law, $\mathrm{d}P(E)/\mathrm{d}\log(E) \propto E^\alpha$ nicely fits the tail in the spectrum towards high energies for $m_{n}>m_{i}$ or $\beta>0.5$~\cite{Devoe2009}. As $\beta$ is increased further, towards even heavier neutrals, the negative exponent $\alpha$ approaches 0, at which point the spectrum does not converge with time anymore and runaway heating starts to dominate the evolution of the ion's energy. The critical mass ratio parameter $\beta_{crit}$ can be found by extrapolating the exponent $\alpha(\beta)$ towards $\alpha(\beta_{crit})=0$.
The quantity $\beta_{crit}$ is not a universal number but is a function of the trap geometry. It depends on the ratio $\frac{\omega_p}{\omega_z}$ or, expressed in the ion trap parameters $a$ and $q$, on $\frac{q^2}{a}=(\frac{2\omega_p}{\omega_z})^2$. The dependency is explained by the fact that axial confinement leads to radial deconfinement and thereby to an increase of the ratio between average micromotion and secular energy. 
For three different cases the extrapolation to $\beta_{crit}$ is shown in \figref{fig:betaCritVsQSqByA}. Our data are compatible with the critical mass ratio previously found for a specific trap geometry~\cite{Devoe2009}. For very elongated traps ($\omega_z\ll\omega_p$) we find $\,\beta_{crit}=0.685$, corresponding to a mass ratio $\frac{	m_{n}}{m_{i}} = 2.17$ .

\subsection{Average energy and lifetime of the ion} 
\begin{figure}[ht]
    \includegraphics{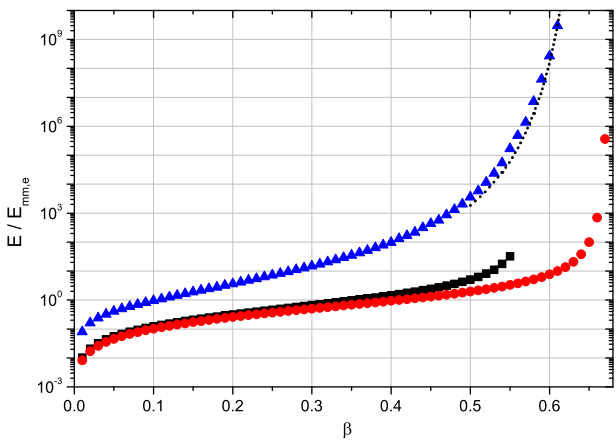}
	\caption[Energy Spectrum Properties vs Beta]{\label{fig:avgEnAndReqTrapDepVsBeta}For different $\beta$ the ion's arithmetic mean energy (black rectangles) and median energy (red circles) are shown in units of $E_{mm,e}$. The arithmetic mean is expected to diverge for $\beta>0.554$. An example for the required trap depth for $P_{loss}<10^{-5}$ is given (blue triangles) and compared to $E_{td} = E_{mm,e}\,\left(P_{loss}\right)^{1/\alpha} $ (dotted line) using $\alpha$ obtained from power law fits to the tails of the energy spectra. All data are for $\frac{q^2}{a}=50$.}
\end{figure}

To evaluate the effectiveness of buffer gas cooling for different mass ratios we calculate the average of the energy spectrum of the ion.
As a physical measure, the arithmetic mean comes close to the definition of a temperature, albeit the clearly non-thermal distribution. We show the arithmetic mean in \figref{fig:avgEnAndReqTrapDepVsBeta} for the case $\frac{q^2}{a}=50$. Although the arithmetic mean diverges for $\alpha\geq-1$, the energy spectrum can still be normalized for $\alpha<0$. In this range the median continues to be a well defined statistical measure for the energy of the ion. 

In experiments, a large probability density at high energies leads to a rapid ion loss due to a finite trap depth $E_{td}$~\cite{Devoe2009,Green2007}. We have numerically evaluated the required trap depth $E_{td}$ to limit the ion loss probability per collision to $P_{loss}$ (\figref{fig:avgEnAndReqTrapDepVsBeta}). For large $\beta$ we can approximate the required trap depth by $E_{td} = E_{mm,e}\,\left(P_{loss}\right)^{1/\alpha} $.
The results show that mass ratios with light neutrals are preferred for efficient buffer gas cooling. However, even for the heavy neutral scenario stable trapping and buffer gas cooling are possible for carefully chosen trap geometry, trap depth and micromotion compensation. 

\subsection{Higher collision rates}
For all the results discussed so far the collision rate has been assumed very low, and as long as the condition $\Gamma \ll \omega$ is fulfilled the results remain unchanged. As the collision rate approaches or even exceeds the trap frequency $\omega_{x,y}$, the probability to have consecutive collisions at correlated positions increases. Under this condition we observe a reduced median energy and an increased power law tail. 

\section{Semiclassical scattering}
\label{sec:secmiclassicalScattering}
Up to this point the classical Langevin model has been used to explore the ion's energy spectrum and its dependence on mass ratio, trap geometry and collision rates. Here we make use of a semiclassical description of the interaction process. The solution to the quantum mechanical scattering problem can be found by expanding the wavefunction into partial waves. 
In the regime where many partial waves contribute, the total elastic cross section scales like $E_c^{-1/3}$~\cite{Massey1934}. The resulting energy dependent scattering rate and angular dependence lead to additional effects as compared to the Langevin model. These are necessary to explain the back action on the neutral cloud.

We model the interaction potential by the long range polarization interaction of \eqrefE{eqn:dxs_C4_potential} plus repulsion at short distances. The full differential cross-section is calculated using~\cite{Sakurai1994} 
\begin{equation}
  \frac{\mathrm{d}\sigma}{\mathrm{d}\Omega}=\frac{1}{k^2} \left| \sum_{l=0}^\infty (2l+1)\,e^{i\eta_l}\,\sin(\eta_l)\,P_l(\cos\theta)\right|^2 \;.
	\label{eqn:dxs_definition1}
\end{equation}
The angular momentum of a partial wave is $\hbar l$ and $\hbar k = \sqrt{2 \mu E_c}$ is the collision momentum. The scattering phase $\eta_l$ can be obtained by solving the radial Schr\"odinger equation which involves the centrifugal potential $\frac{\hbar^2\,l(l+1)}{2 \mu R^2}$. The resulting centrifugal barrier increases in height with angular momentum ($\sim(\hbar l)^4)$. Partial waves with $l<l_0=1/\hbar \sqrt{2\mu\sqrt{2 C_4 E_c}}$ have a collision energy larger than the height of the centrifugal barrier, probe the deep potential well and are reflected from the hard core. The exact form of the potential, relevant to determine $\eta_l$, is typically not known. Therefore phase shifts $\eta_l$ for $l<l_0$ are assumed to be uniformly distributed within $[0,2\pi)$~\cite{Cote2000}. In this approximation each partial wave contributes with $\sigma_l=\frac{2 \pi l}{k^2}$ to the total cross-section. Summing $\sigma_l$ up to $l_0$ reproduces the Langevin cross-section. 

The full quantum mechanical cross-section includes additional contributions from partial waves with $l>l_0$. As the centrifugal barrier is higher than the collision energy, these partial waves are scattered from the centrifugal barrier, if tunnelling effects are neglected. The phase shifts can be semiclassically approximated by~\cite{Cote2000,Massey1934}
\begin{equation}
  \eta_l = -\frac{\mu}{\hbar^2}\int_{R_0}^\infty{\frac{V(R)}{\sqrt{k^2-\frac{(l+1/2)^2}{R^2}}}\mathrm{d}R}
	\label{eqn:dxs_formula_eta_l}
\end{equation} 
with $R_0 = \frac{l+1/2}{k}$. 

\subsection{Modeling the differential cross-section}
\label{sec:diffXSModel}
\begin{figure}[ht]
    \includegraphics{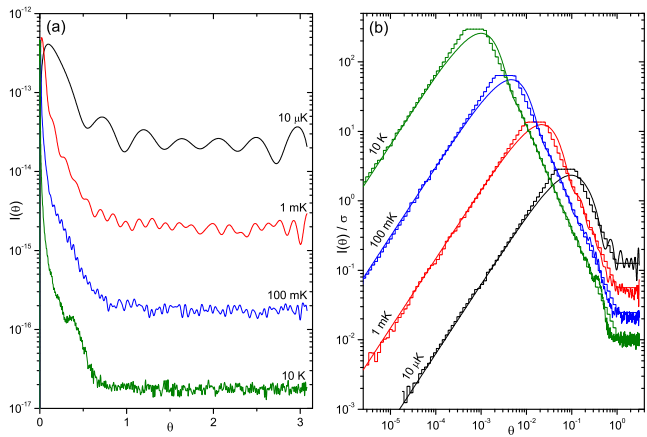}
	\caption[Numerically obtained $I(\theta,E_c)$ for ${}^{174}\text{Yb}-{}^{87}\text{Rb}$]{\label{fig:dxs_I_th}(a) The probability distributions $I(\theta,E_c)$ for the scattering angle $\theta$ in elastic collisions between ${}^{174}\text{Yb}^+$ and ${}^{87}\text{Rb}$ are numerically calculated. The unknown scattering phases for close encounters ($l<l_0$) are chosen randomly. The four curves are for different collision energies $E_c$. 
The forward scattering peak at small $\theta$ is more pronounced for higher energies. Its shape is emphasized in (b) where the scattering angle $\theta$ is displayed logarithmically. 
\label{fig:dxsITthLog_vs_randomThetaGen}The normalized $I(\theta,E_c)$ (smooth lines) for the four different energies are compared with logarithmically binned output from the random $\theta$-generating function (step like lines). The approximation is optimized to reproduce the height and position of the forward scattering peak. }
\end{figure}

The probability distribution for the deflection angle $\theta$
\begin{equation}
  I(\theta,E_c)=\frac{\mathrm{d}\sigma}{\mathrm{d}\theta}= \int_0^{2\pi}\frac{\mathrm{d}\sigma}{\mathrm{d}\Omega}\sin\theta\,\mathrm{d}\phi
  \label{eqn:dxs_dSigmaBydTheta}
\end{equation} 
is numerically calculated using \eqrefE{eqn:dxs_definition1}, \eqrefE{eqn:dxs_formula_eta_l} and randomly distributed $\eta_l$ for $l<l_0$. We sum \eqrefE{eqn:dxs_definition1} for $l$ up to 20000 and average over 100 different random sets of $\eta_l$. The differential cross-section calculated in this way depends on the reduced mass $\mu$, the collision energy $E_c=\frac{\hbar^2\,k^2}{2\,\mu}$ and $C_4$ only. For the case of a ${}^{174}\text{Yb}^+$ ion colliding with a ${}^{87}\text{Rb}$ atom, four probability distributions of the form \eqrefE{eqn:dxs_dSigmaBydTheta} are shown in \figref{fig:dxs_I_th}. The main feature of $I(\theta,E_c)$ is a forward scattering peak, which gets more pronounced as the energy increases~\cite{Zhang2009}. The integral of the differential cross-section reproduces the expected $E_c^{-1/3}$ energy dependence, and its magnitude is in agreement with~\cite{Cote2000}. 

To implement the differential cross-section in the Monte Carlo simulation a parameterization of the normalized $I(\theta,E_c)$ is used to create a function that returns a random $\theta$ for a given collison energy. 
The distribution $I(\theta,E_c)$ is modelled in four intervals using two power laws ($\propto \theta^{p}$), a flat top and a flat background. The parameters peak height, background offset and interval limits are energy dependent and are well approximated by power laws ($\propto \,E_c^{p'}$). These power laws are obtained from fits to differential cross-sections for more than 30 different energies $E_c$ in the range between $k_B\times1\,\mu \text{K}$ and $k_B\times100\,\text{K}$. The $\theta$-generating function uses these parameter functions and inverse transform sampling. \figref{fig:dxsITthLog_vs_randomThetaGen}b compares sampled output of the $\theta$-generating function with the normalized differential cross-sections. 

\subsection{Effects of the energy dependent scattering rate on the ion energy spectrum}
\label{sec:effOnIonSpectrum}
We have simulated the kinetics of the ion in an ultracold buffer gas using the parametrized differential cross-section to investigate how this affects the ion energy spectrum.
Different from the Langevin case, the collision rate depends on the instantaneous ion energy. Therefore efficient simulation relies on the collision time sampling described in Appendix~\ref{app:collTimeSampling}. 
The ion energies are binned and weighted by the time the ion remains at the specific energy. 

We have performed simulations for an ion trap with frequencies $\omega_{x,y,z}=2\pi\times\{151,153,42\}\, \text{kHz},\,\Omega_T=2\pi\times42.5\,\text{MHz}$, excess micromotion parameters $\Delta x=\Delta y=2\,\mu \text{m}$ and neutral density $n=10^{18}\,\text{m}^{-3}$ for the system ${}^{174}\text{Yb}^+$ - ${}^{87}\text{Rb}$, reproducing conditions comparable to~\cite{Zipkes2010b}.
We find good agreement with the previous results from the Langevin model and conclude that the Langevin model is sufficient to describe the ion's energy statistics. Formally, however, the system does not necessarily scale only with the excess-micromotion energy $E_{mm,e}$ anymore, as the differential cross-section introduces its own energy scale. 
In the next section we will demonstrate that the full differential cross-section is necessary to predict effects on the cold neutral atoms.
 
\subsection{Neutral cloud evolution}
\label{sec:neutrCloud}
Ultracold neutral atomic clouds have atom numbers ranging up to $10^9$. Compared to a room temperature buffer gas, this limited number of atoms and the good isolation from the environment allow the observation of collision effects on the neutral gas. The main observables are the number of neutral atoms $N_a$ and their temperature $T_a$. In experiments, these values can be obtained from time-of-flight imaging and are suitable to verify the simulation model. 

\begin{figure}[ht]
    \includegraphics{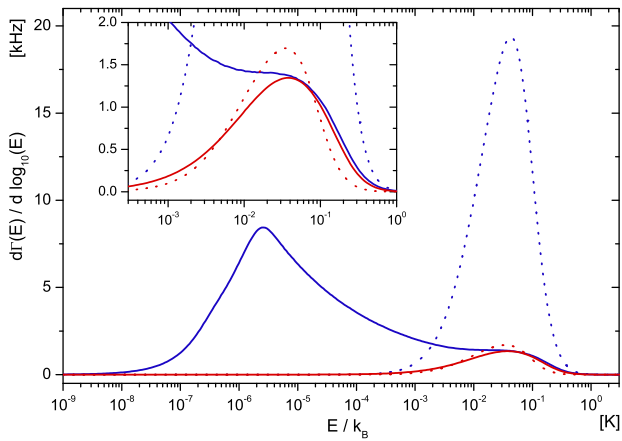}
	\caption[Collision and transferred energy spectrum]{\label{fig:collAndTransEnSpect} Distributions of collision energy $E_c$ (dotted line) and transferred energy $E_t$ (solid line). The data are obtained by binning $10^8$ collisions on a logarithmic energy scale, comparing Langevin scattering (red) with the full semiclassical differential cross-section (blue). The vertical axis indicates the scattering rate in $\text{kHz}$ per decade of energy at which collisions with the specific energies occur. The inset is a magnification of the region relevant for Langevin scattering. The settings for this simulation run are  $n=10^{18}\,\text{m}^{-3}$ for the uniform neutral atom density and $\Delta x=\Delta y=2\,\mu \text{m}$, resulting in an excess-micromotion energy of $E_{mm,e}/k_B = 160\,\text{mK}$. Note the similarity between the two models for large and the significant difference for small transferred energies $E_t$. }
\end{figure}

The back-action of the ion onto the neutral gas is a result of the energy transfer per collision.  
For general two-body elastic collisions it is given by 
\begin{equation}
  E_t = 4\,(1-\beta)\,E_c\,\sin^2(\theta/2)\;,
	\label{eqn:colETrans}
\end{equation} 
depending on the scattering angle $\theta$. 
For very small deflections $\theta \ll 1$, resulting from the forward scattering peak in the differential cross-section, only very little energy is transferred to the neutral atom. 
The distribution of transferred energies $E_t$, shown in \figref{fig:collAndTransEnSpect}, can be understood as a convolution of the collision energy distribution and the energy dependent differential cross-section. 

\figref{fig:collAndTransEnSpect} also shows distributions for $E_c$ and $E_t$ obtained using the Langevin model for comparison. The Langevin collision rate is obviously smaller, explained by the different energy dependence of the cross-sections, $\sigma_L\propto E_c^{-1/2}$ vs $\sigma\propto E_c^{-1/3}$. The semiclassical distribution of $E_c$ is also slightly shifted towards higher energies with respect to the Langevin $E_c$, since collisions are more likely to happen at higher energies. The transferred energies differ only little between to two models for $E_t \gtrsim k_B\times0.03\,\text{K}$. This reflects the statement that the ion's mobility is well described by Langevin type collisions. However, the significant peak at low $E_t$ of the semiclassical distribution causes most of the effects on the cold neutral gas. 

So far all the results were obtained assuming a uniform density distribution of neutral atoms. This will now be replaced by a spatial distribution for a thermal gas in a harmonic trap with temperature $T_a$ and atom number $N_a$.
Considering a finite trap depth $E_{td,a}$ for the neutral atoms, every collision with $E_t> E_{td,a}$ will lead to an atom loss, whereas every $E_t< E_{td,a}$ will increase the temperature $T_a$. The simplified model used in the simulation assumes immediate thermal equilibration of the neutral gas. 
Then a collision with $E_t< E_{td,a}$ simply increases $T_a$ by $\frac{E_t}{3\,k_B\,N_a}$. The loss of an atom for $E_t> E_{td,a}$ will decrease $N_a$ by $1$ but also affect the temperature depending on the atoms energy. 
The new temperature is calculated 
\begin{equation}
  T_{a,f} = \frac{N_{a,i}\,3\,k_B\,T_{a,i} - \frac{3}{2}\,k_B\,T_{a,i} - E_{pot}(\vec{r})}{(N_{a,i}-1)\,3\,k_B}\;
	\label{eqn:neutEChangeByLoss}
\end{equation}
using the total energy of the neutral cloud, and the potential and average kinetic energy of the lost atom known from the position $\vec{r}$ of the collision. This can lead to evaporative cooling or heating effects depending on the position of the ion in the neutral gas. 

For ion trajectories larger than the size of the buffer gas the ion can only collide in the centre of the trap where its motional energy is mostly related to the secular motion rather than the micromotion. This suppresses the power law tail of the ion energy distribution and reduces the ion's average energy.
Hence, using tight traps to confine the neutral atoms might help to overcome the constraints on ion trap depth, trap geometry and mass ratio.

\subsection{Comparison to experimental data}
\label{sec:expComp}
\begin{figure}[ht]
    \includegraphics{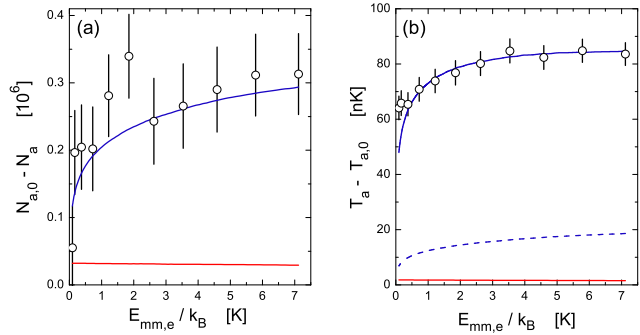}
	\caption[Comparison to Experimental Data]{\label{fig:expComp} Comparison between experimental measurements (black circles) and the simulation predictions for neutral atom loss (a) and the neutral temperature increase (b). The semiclassical model (blue) fits the data well. The Langevin model (red) systematically underestimates the collision effects on the neutral atoms. The contribution of evaporative heating (\eqrefE{eqn:neutEChangeByLoss}) in the semiclassical model is indicated with the blue dashed line. The experimental data are taken from~\cite{Zipkes2010b}.}
\end{figure}
Here we compare the simulation predictions for the effects on neutral atoms with experimental data from $\text{Yb}^+ - \text{Rb}$~\cite{Zipkes2010b}. The measured quantities are the loss of neutral atoms and temperature increase of a cold thermal cloud, after $8\,\text{s}$ of interaction and for different excess-micromotion energies $E_{mm,e}$. \figref{fig:expComp} displays the data together with the simulation results. 

Initial conditions for the neutral ${}^{87}\text{Rb}$ cloud in the simulation are given by $T_{a,0} = 250\,\text{nK}$ and $N_{a,0} = 2.25\times 10^6$. Neutral trap frequencies of $2\pi\times\{28,28,8\}\,\text{Hz}$ result in an initial central density of $n(0)=1.9\times 10^{18}\,\text{m}^{-3}$. The neutral trap depth is $E_{td,a} = k_B\times8\,\mu\text{K}$.  

A single ${}^{172}\text{Yb}^+$ ion is trapped with parameters given in section~\ref{sec:effOnIonSpectrum}. The excess micromotion parameters $\Delta x$ and $\Delta y$ are varied between $0$ and $15\,\mu\text{m}$, and additionally there is micromotion along the trap symmetry axis with $c_z = 2\,\mu\text{m}$. The excess-micromotion energy scale, $E_{mm,e}/k_B$ thus ranges from $90\,\text{mK}$ to $7\,\text{K}$.

The semiclassical simulation predicts both the shape and the magnitude of the experimental results well. In contrast, the Langevin scattering is not suited to model these measurements because the ultracold atoms are highly sensitive to small energy transfer $E_t$. They correspond to collisions with small deflection angles, which are neglected by the Langevin model, cf.~\figref{fig:collAndTransEnSpect}.

\section{Conclusion}
We have numerically investigated the kinetics of a single trapped ion interacting with an ultracold neutral gas. 
Our results explain the effects of the mass ratio, trap geometry and excess micromotion on the ion's energy spectrum.
We have applied two different collision models of the atom-ion interaction, the Langevin and the semiclassical scattering model. 
Both yield similar ion energy statistics.
The greater simplicity of the Langevin model introduces a characteristic energy scale for the ion energy statistics.
However, experimentally observed effects on the neutral cloud can only be explained by the semiclassical model. Forward scattering events with small energy transfer are affecting both the neutral cloud temperature and the atom loss rate. 
Cold atom-ion collisions could then be used to locally remove atoms resulting in efficient cooling of quantum gases.

\begin{acknowledgments}
We thank EPSRC (EP/F016379/1, EP/H005676/1), ERC (Grant No. 240335), and the Herchel Smith Fund for support.
\end{acknowledgments}

\appendix

\section{Collision time sampling}
\label{app:collTimeSampling}
A scattering rate $\Gamma$ sets the probability for a collision to take place within a time interval $\mathrm{d}t$. In our case 
\begin{equation}
	\Gamma(t) = n(\vec{x})\,\sigma(E_c)\,v_{ion}(t) 	\;
	\label{eqn:scattRate}
\end{equation}
with $n$ being the neutral atom density, $\sigma$ the cross-section and the velocity $v_{ion}$ as defined in \eqrefE{eqn:ionMotionVelocity1}. Usually the density is a function on the ion's position, the cross-section depends on the collision energy and $v_{ion}$ oscillates rapidly in time, leading to a $\Gamma(t)$ with non-trivial time dependence. In the following we explain the method used to randomly generate collision times $t$ with the exact distribution defined by $\Gamma(t)$. 

In general, the process of an event (collision) taking place with a rate $\Gamma(t)$ can be modelled using a differential equation for the probability $Q(t)$ that the event has not yet happened after the time $t$,
\begin{equation}
	\mathrm{d}Q(t) = -\Gamma(t)\,Q(t)\,\mathrm{d}t 	\;.
	\label{eqn:diffEqnEventWithRate}
\end{equation} 
The probability distribution for an event to take place after the time $t$ is defined by $P(t)=-\mathrm{d}Q/\mathrm{d}t$. 
In the simple case with constant $\Gamma(t)=\Gamma_0$ the solution is
\begin{equation}
	P_{\Gamma_0}(t)=\Gamma_0\,\exp(-\Gamma_0\,t) 	\;
	\label{eqn:constantGamma_randT_pdf}
\end{equation} 
and a random time can be obtained using inverse transform sampling, 
\begin{equation}
	t=-1/\Gamma_0\,\log(r) 	\;
	\label{eqn:constantGamma_randT_its}
\end{equation} 
with $r$ being a uniformly distributed random number in the interval $(0,1]$. For time dependent $\Gamma(t)$ the analytic solution for the probability distribution function is 
\begin{equation}
	P(t)=\Gamma(t)\,\exp\Big(-\int_0^t\Gamma(t_1)\,\mathrm{d}t_1\Big) 	\;.
	\label{eqn:varGamma_randT_pdf}
\end{equation} 
Non-trivial time dependence of $\Gamma(t)$ usually requires a numerical approach to sample $t$ from \eqrefE{eqn:varGamma_randT_pdf}.
One straight-forward method would be to discretize time into small steps, calculate $\Gamma(t)$ and its contribution to $P(t)$ for every step, thereby numerically integrating the function $P(t)$ up to a randomly chosen trigger value, at which point the event takes place.

However, another method is employed here, which proves to be much faster and does not suffer from discretization errors. It works for $\Gamma(t)$ that have an upper bound $\Gamma_m=\sup(\Gamma(t))$, or where such a condition can be enforced by introducing a cutoff. In the specific case of the trapped ion, all factors in \eqrefE{eqn:scattRate} are easily limited by considering the energies $E_j$ defining the trajectory and the excess micromotion parameters and using the peak neutral density. This upper bound can be adjusted after each collision event having affected $E_j$ and $n(\vec{x})$.

The algorithm works by advancing the system by a time $t$ according to \eqrefE{eqn:constantGamma_randT_its} with $\Gamma_0=\Gamma_m$. Then the rescaled rate
\begin{equation}
 \gamma(t)=\Gamma(t)/\Gamma_m
 \label{eqn:varGamma_randT_GammaRescale}
\end{equation} 
  is calculated for the resulting state of the system after the time $t$ and an event takes place if $\gamma(t)>r$, with $r$ being another uniformly distributed random number in the interval $[0,1)$. If the event does not take place ($\gamma(t)\le r$) the algorithm iteratively loops back to advance the system by an additional $t$, again according to \eqrefE{eqn:constantGamma_randT_its}.
The method is exact in that it reproduces the probability distribution function \eqrefE{eqn:varGamma_randT_pdf}. A proof follows below. 
The efficiency of the method is the ratio of the average of $\Gamma(t)$ and $\Gamma_m$,  $\epsilon=\langle\Gamma(t)\rangle/\Gamma_m=\langle\gamma(t)\rangle$. This means that in order to generate $N$ events it can be expected that $\Gamma(t)$ needs to be evaluated $N/\epsilon$ times.

\textit{Proof:} 
We start from writing an expression for the probability distribution function $P_s(t)$ obtained with the suggested method. Since the final $t$ can be the result of any number of iterations, $P_s(t)$ is a sum of all these possibilities, $P_s(t) = P_{s,1} + P_{s,2} + P_{s,3} + \ldots $, where $P_{s,i}$ is the probability that $t$ results as the time of event after $i$ iterations. The first few terms are given below.
\begin{equation}
	P_{s,1}=\gamma(t)\,P_{\Gamma_m}(t) 	\;
	\label{eqn:varGamma_randT_prove_p1}
\end{equation} 
\begin{equation}
	P_{s,2}=\gamma(t)\,\int_0^t P_{\Gamma_m}(t_1)\,P_{\Gamma_m}(t-t_1)\,\big(1-\gamma(t1)\big)\,\mathrm{d}t_1 	\;
	\label{eqn:varGamma_randT_prove_p2}
\end{equation} 
\begin{multline}
	P_{s,3}=\gamma(t)\,\int_0^t \int_0^{t_1} P_{\Gamma_m}(t_2)\,P_{\Gamma_m}(t_1-t_2)\, \\ P_{\Gamma_m}(t-t_1)\,\big(1-\gamma(t_1)\big)\,\big(1-\gamma(t_2)\big)\,\mathrm{d}t_2\,\mathrm{d}t_1 	\;
	\label{eqn:varGamma_randT_prove_p3}
\end{multline} 
Note that the products of $P_{\Gamma_m}$ (as defined in \eqrefE{eqn:constantGamma_randT_pdf}) in the integrals always combine to $\Gamma_m^{(i-1)}P_{\Gamma_m}(t)$. 
Therefore $P_{\Gamma_m}(t)$ is taken out of the sum as a common prefactor,
\begin{multline}
	P_s(t)=\gamma(t)\,P_{\Gamma_m}(t) \Big(1 + \Gamma_m \int_0^t \big(1-\gamma(t1)\big)\,\mathrm{d}t_1  \\
	 + {\Gamma_m}^2 \int_0^t \int_0^{t_1} \big(1-\gamma(t_1)\big)\,\big(1-\gamma(t_2)\big)\,\mathrm{d}t_2\,\mathrm{d}t_1 +\ldots \Big) 	\; .
	 \label{eqn:varGamma_randT_prove_ps_2}
 \end{multline}
Now the upper boundaries of all the partial integrals can be set equal to $t$, since they only induce ordering to the time series $\{t_1,t_2,\ldots,t_n\}$. This rescales the terms by the number of possible orderings ($n!$). Then the partial integrals reduce to a single integral to the power of $n$,
\begin{multline}
	 \int_0^t \int_0^{t_1} \ldots \int_0^{t_{n-1}} \big(1-\gamma(t_1)\big)\,\big(1-\gamma(t_2)\big)	\ldots  \\ \ldots \big(1-\gamma(t_n)\big) \, \mathrm{d}t_n\ldots\,\mathrm{d}t_2\,\mathrm{d}t_1 \\
	 = \frac{1}{n!}  \int_0^t \int_0^t \ldots \int_0^t \big(1-\gamma(t_1)\big)\,\big(1-\gamma(t_2)\big)	\ldots \\ \ldots \big(1-\gamma(t_n)\big) \, \mathrm{d}t_n\ldots\,\mathrm{d}t_2\,\mathrm{d}t_1 \\
	 = \frac{1}{n!} \Big( \int_0^t \big(1-\gamma(t_1)\big)\,\mathrm{d}t_1 \Big)^n \; .
	\label{eqn:varGamma_randT_prove_ordering}
\end{multline} 
Combining this with \eqrefE{eqn:varGamma_randT_prove_ps_2} gives
\begin{multline}
	P_s(t) = \gamma(t)\,P_{\Gamma_m}(t)\sum_{n=0}^\infty \frac{{\Gamma_m}^n}{n!} \Big( \int_0^t \big(1-\gamma(t_1)\big)\,\mathrm{d}t_1 \Big)^n \\
	 = \gamma(t)\,P_{\Gamma_m}(t)\exp\Big( \Gamma_m\int_0^t \big(1-\gamma(t_1)\big)\,\mathrm{d}t_1 \Big)  \;,
	\label{eqn:varGamma_randT_prove_ps_3}
\end{multline} 
and it follows with \eqrefE{eqn:constantGamma_randT_pdf} and \eqrefE{eqn:varGamma_randT_GammaRescale}
\begin{multline}
	P_s(t) = \gamma(t)\,\Gamma_m\exp\Big(-\Gamma_m t +\Gamma_m\int_0^t \big(1-\gamma(t_1)\big)\,\mathrm{d}t_1 \Big) \\
	= \Gamma(t) \exp\Big(-\int_0^t \Gamma(t_1)\,\mathrm{d}t_1 \Big)  \; .
	\label{eqn:varGamma_randT_prove_ps_4}
\end{multline} 
This is identical to \eqrefE{eqn:varGamma_randT_pdf} which proves that the method reproduces the exact probability distribution.

\section{Structure of the simulation loop}
\label{app:simulationLoop}
The following is a short description of how the main simulation loop has been implemented.
\texttt{
\begin{itemize}
  \item[1)] setup system configuration: $\omega_j$, $\Omega_T$ and excess micromotion, $\beta$, neutral trap frequencies, $T_a$, $N_a$, $n(\vec{x})$, $\frac{\mathrm{d}\sigma}{\mathrm{d}\Omega}$, ...
  \item[2)] select initial state $( E_j, \varphi_j )_{j\in\{x,y,z\}} $
  \item[3)] calculate maximal collision rate $\Gamma_m$, considering $n_{max}(T_a, N_a)$ and $E_j$
  \item[4)] increment time $t$ by \eqrefE{eqn:constantGamma_randT_its}
  \item[5)] calculate ion position and velocity, 
    \\using \eqrefE{eqn:posIonTa},\eqref{eqn:ionMicromotionVelocity1},\eqref{eqn:ionMotionVelocity1}, 
    $( E_j, \varphi_j, t ) \rightarrow (\vec{r}_{ion}, \vec{v}_{ion})$ 
  \item[6)] calculate $\gamma(t)$ with \eqrefE{eqn:varGamma_randT_GammaRescale}, \\ proceed to 7) with probability $\gamma(t)$, \\else go back to 4)
  \item[7)] choose collision parameters ($\theta$, $\phi$) according to $\frac{\mathrm{d}\sigma}{\mathrm{d}\Omega}$
  \item[8)] update neutral atom parameters $T_a$, $N_a$, \\using \eqrefE{eqn:colETrans},\eqref{eqn:neutEChangeByLoss} and $E_{td,a}$
  \item[9)] apply scattering \eqrefE{eqn:mcDeflection} to $\vec{v}_{ion}$
  \item[10)] calculate new trajectory parameters, 
    \\using \eqrefE{eqn:posIonTa},\eqref{eqn:ionMicromotionVelocity1},\eqref{eqn:ionMotionVelocity1}, 
    $( \vec{r}_{ion}, \vec{v}_{ion}, t ) \rightarrow (E_j, \varphi_j)$ 
  \item[11)] loop back to 3)
\end{itemize}
}
The setup of the system configuration in \texttt{1)} contains mainly parameters which do not change during the collisions, such as trap frequencies or the atom-ion mass ratio. Exceptions are the neutral atom number $N_a$ and temperature $T_a$, which act as initial conditions. The initial state for the ion energy in \texttt{2)} is mostly unimportant, as the simulation will iterate over many collisions and the information of the initial state is lost after a few collisions. When looking at steady state statistics of the ion energy, the values after the first few collisions can simply be ignored, thus effectively letting the system evolve for a short time before measuring its properties. 
The points \texttt{3)} to \texttt{7)} implement the collision time sampling algorithm described in Appendix\,\ref{app:collTimeSampling}. The maximally possible collision rate $\Gamma_m$ is calculated from the peak density $n_{max}(T_a, N_a)$ of the neutral atoms and on the maximum value of $\sigma(E_c)\,v_{ion}(t)$. The latter is limited by the highest possible ion velocity, which depends on the secular energy $E_j$ and the excess micromotion. 
The calculation of $\gamma(t)$ in \texttt{6)} uses the $\vec{v}_{ion}$ and $\vec{r}_{ion}$ obtained in \texttt{5)} to determine $\Gamma(t)$ with \eqrefE{eqn:scatteringRate}. If a collision takes place at the chosen time $t$, \texttt{7)} gives the scattering angles according to the differential cross-section. \texttt{8)} simulates the back-action on the neutral atoms. \texttt{9)} modifies the velocity of the ion. In \texttt{10)} the new trajectory parameters $E_j$ and $\varphi_j$ are determined, to represent the motion of the ion up to the next collision. 

Information on any system parameter can be retrieved at user-defined points within the simulation loop. The ion energy is typically registered after \texttt{10)}, the collision and transferred energies, $E_c$ and $E_t$ after \texttt{8)}. Random sampling is used in \texttt{4)} and \texttt{6)}, to determine the time of collision, and in \texttt{7)}, for the scattering angles.


\begin{thebibliography}{46}%
\makeatletter
\providecommand \@ifxundefined [1]{%
 \@ifx{#1\undefined}
}%
\providecommand \@ifnum [1]{%
 \ifnum #1\expandafter \@firstoftwo
 \else \expandafter \@secondoftwo
 \fi
}%
\providecommand \@ifx [1]{%
 \ifx #1\expandafter \@firstoftwo
 \else \expandafter \@secondoftwo
 \fi
}%
\providecommand \natexlab [1]{#1}%
\providecommand \enquote  [1]{``#1''}%
\providecommand \bibnamefont  [1]{#1}%
\providecommand \bibfnamefont [1]{#1}%
\providecommand \citenamefont [1]{#1}%
\providecommand \href@noop [0]{\@secondoftwo}%
\providecommand \href [0]{\begingroup \@sanitize@url \@href}%
\providecommand \@href[1]{\@@startlink{#1}\@@href}%
\providecommand \@@href[1]{\endgroup#1\@@endlink}%
\providecommand \@sanitize@url [0]{\catcode `\\12\catcode `\$12\catcode
  `\&12\catcode `\#12\catcode `\^12\catcode `\_12\catcode `\%12\relax}%
\providecommand \@@startlink[1]{}%
\providecommand \@@endlink[0]{}%
\providecommand \url  [0]{\begingroup\@sanitize@url \@url }%
\providecommand \@url [1]{\endgroup\@href {#1}{\urlprefix }}%
\providecommand \urlprefix  [0]{URL }%
\providecommand \Eprint [0]{\href }%
\providecommand \doibase [0]{http://dx.doi.org/}%
\providecommand \selectlanguage [0]{\@gobble}%
\providecommand \bibinfo  [0]{\@secondoftwo}%
\providecommand \bibfield  [0]{\@secondoftwo}%
\providecommand \translation [1]{[#1]}%
\providecommand \BibitemOpen [0]{}%
\providecommand \bibitemStop [0]{}%
\providecommand \bibitemNoStop [0]{.\EOS\space}%
\providecommand \EOS [0]{\spacefactor3000\relax}%
\providecommand \BibitemShut  [1]{\csname bibitem#1\endcsname}%
\let\auto@bib@innerbib\@empty
\bibitem [{\citenamefont {Cirac}\ and\ \citenamefont
  {Zoller}(1995)}]{Cirac1995}%
  \BibitemOpen
  \bibfield  {author} {\bibinfo {author} {\bibfnamefont {J.~I.}\ \bibnamefont
  {Cirac}}\ and\ \bibinfo {author} {\bibfnamefont {P.}~\bibnamefont {Zoller}},\
  }\href {\doibase 10.1103/PhysRevLett.74.4091} {\bibfield  {journal} {\bibinfo
   {journal} {Phys. Rev. Lett.}\ }\textbf {\bibinfo {volume} {74}},\ \bibinfo
  {pages} {4091} (\bibinfo {year} {1995})}\BibitemShut {NoStop}%
\bibitem [{\citenamefont {Blatt}\ and\ \citenamefont
  {Wineland}(2008)}]{Blatt2008}%
  \BibitemOpen
  \bibfield  {author} {\bibinfo {author} {\bibfnamefont {R.}~\bibnamefont
  {Blatt}}\ and\ \bibinfo {author} {\bibfnamefont {D.~J.}\ \bibnamefont
  {Wineland}},\ }\href@noop {} {\bibfield  {journal} {\bibinfo  {journal}
  {Nature}\ }\textbf {\bibinfo {volume} {453}},\ \bibinfo {pages} {1008}
  (\bibinfo {year} {2008})}\BibitemShut {NoStop}%
\bibitem [{\citenamefont {Blatt}\ \emph {et~al.}(1992)\citenamefont {Blatt},
  \citenamefont {Gill},\ and\ \citenamefont {Thompson}}]{Blatt1992}%
  \BibitemOpen
  \bibfield  {author} {\bibinfo {author} {\bibfnamefont {R.}~\bibnamefont
  {Blatt}}, \bibinfo {author} {\bibfnamefont {P.}~\bibnamefont {Gill}}, \ and\
  \bibinfo {author} {\bibfnamefont {R.}~\bibnamefont {Thompson}},\ }\href
  {http://www.ingentaconnect.com/content/tandf/tmop/1992/00000039/00000002/art%
00001} {\bibfield  {journal} {\bibinfo  {journal} {Journal of Modern Optics}\
  }\textbf {\bibinfo {volume} {39}},\ \bibinfo {pages} {193} (\bibinfo {year}
  {1992})}\BibitemShut {NoStop}%
\bibitem [{\citenamefont {Krems}(2008)}]{Krems2008}%
  \BibitemOpen
  \bibfield  {author} {\bibinfo {author} {\bibfnamefont {R.~V.}\ \bibnamefont
  {Krems}},\ }\href@noop {} {\bibfield  {journal} {\bibinfo  {journal} {Phys.
  Chem. Chem. Phys.}\ }\textbf {\bibinfo {volume} {10}},\ \bibinfo {pages}
  {4079} (\bibinfo {year} {2008})}\BibitemShut {NoStop}%
\bibitem [{\citenamefont {Cetina}\ \emph {et~al.}(2007)\citenamefont {Cetina},
  \citenamefont {Grier}, \citenamefont {Campbell}, \citenamefont {Chuang},\
  and\ \citenamefont {Vuleti\ifmmode~\acute{c}\else \'{c}\fi{}}}]{Cetina2007}%
  \BibitemOpen
  \bibfield  {author} {\bibinfo {author} {\bibfnamefont {M.}~\bibnamefont
  {Cetina}}, \bibinfo {author} {\bibfnamefont {A.}~\bibnamefont {Grier}},
  \bibinfo {author} {\bibfnamefont {J.}~\bibnamefont {Campbell}}, \bibinfo
  {author} {\bibfnamefont {I.}~\bibnamefont {Chuang}}, \ and\ \bibinfo {author}
  {\bibfnamefont {V.}~\bibnamefont {Vuleti\ifmmode~\acute{c}\else
  \'{c}\fi{}}},\ }\href {\doibase 10.1103/PhysRevA.76.041401} {\bibfield
  {journal} {\bibinfo  {journal} {Phys. Rev. A}\ }\textbf {\bibinfo {volume}
  {76}},\ \bibinfo {pages} {041401} (\bibinfo {year} {2007})}\BibitemShut
  {NoStop}%
\bibitem [{\citenamefont {Grier}\ \emph {et~al.}(2009)\citenamefont {Grier},
  \citenamefont {Cetina}, \citenamefont {Oru\v{c}evi\'{c}},\ and\ \citenamefont
  {Vuleti\'{c}}}]{Grier2009}%
  \BibitemOpen
  \bibfield  {author} {\bibinfo {author} {\bibfnamefont {A.~T.}\ \bibnamefont
  {Grier}}, \bibinfo {author} {\bibfnamefont {M.}~\bibnamefont {Cetina}},
  \bibinfo {author} {\bibfnamefont {F.}~\bibnamefont {Oru\v{c}evi\'{c}}}, \
  and\ \bibinfo {author} {\bibfnamefont {V.}~\bibnamefont {Vuleti\'{c}}},\
  }\href {\doibase 10.1103/PhysRevLett.102.223201} {\bibfield  {journal}
  {\bibinfo  {journal} {Phys. Rev. Lett.}\ }\textbf {\bibinfo {volume} {102}},\
  \bibinfo {eid} {223201} (\bibinfo {year} {2009})}\BibitemShut {NoStop}%
\bibitem [{\citenamefont {Zipkes}\ \emph
  {et~al.}(2010{\natexlab{a}})\citenamefont {Zipkes}, \citenamefont {Palzer},
  \citenamefont {Sias},\ and\ \citenamefont {K{\"o}hl}}]{Zipkes2010}%
  \BibitemOpen
  \bibfield  {author} {\bibinfo {author} {\bibfnamefont {C.}~\bibnamefont
  {Zipkes}}, \bibinfo {author} {\bibfnamefont {S.}~\bibnamefont {Palzer}},
  \bibinfo {author} {\bibfnamefont {C.}~\bibnamefont {Sias}}, \ and\ \bibinfo
  {author} {\bibfnamefont {M.}~\bibnamefont {K{\"o}hl}},\ }\href@noop {}
  {\bibfield  {journal} {\bibinfo  {journal} {Nature}\ }\textbf {\bibinfo
  {volume} {464}},\ \bibinfo {pages} {388} (\bibinfo {year}
  {2010}{\natexlab{a}})}\BibitemShut {NoStop}%
\bibitem [{\citenamefont {Zipkes}\ \emph
  {et~al.}(2010{\natexlab{b}})\citenamefont {Zipkes}, \citenamefont {Palzer},
  \citenamefont {Ratschbacher}, \citenamefont {Sias},\ and\ \citenamefont
  {K{\"o}hl}}]{Zipkes2010b}%
  \BibitemOpen
  \bibfield  {author} {\bibinfo {author} {\bibfnamefont {C.}~\bibnamefont
  {Zipkes}}, \bibinfo {author} {\bibfnamefont {S.}~\bibnamefont {Palzer}},
  \bibinfo {author} {\bibfnamefont {L.}~\bibnamefont {Ratschbacher}}, \bibinfo
  {author} {\bibfnamefont {C.}~\bibnamefont {Sias}}, \ and\ \bibinfo {author}
  {\bibfnamefont {M.}~\bibnamefont {K{\"o}hl}},\ }\href {\doibase
  10.1103/PhysRevLett.105.133201} {\bibfield  {journal} {\bibinfo  {journal}
  {Phys. Rev. Lett.}\ }\textbf {\bibinfo {volume} {105}},\ \bibinfo {pages}
  {133201} (\bibinfo {year} {2010}{\natexlab{b}})}\BibitemShut {NoStop}%
\bibitem [{\citenamefont {Schmid}\ \emph {et~al.}(2010)\citenamefont {Schmid},
  \citenamefont {H\"arter},\ and\ \citenamefont
  {Hecker-Denschlag}}]{Schmid2010}%
  \BibitemOpen
  \bibfield  {author} {\bibinfo {author} {\bibfnamefont {S.}~\bibnamefont
  {Schmid}}, \bibinfo {author} {\bibfnamefont {A.}~\bibnamefont {H\"arter}}, \
  and\ \bibinfo {author} {\bibfnamefont {J.}~\bibnamefont {Hecker-Denschlag}},\
  }\href {\doibase 10.1103/PhysRevLett.105.133202} {\bibfield  {journal}
  {\bibinfo  {journal} {Phys. Rev. Lett.}\ }\textbf {\bibinfo {volume} {105}},\
  \bibinfo {pages} {133202} (\bibinfo {year} {2010})}\BibitemShut {NoStop}%
\bibitem [{\citenamefont {Ravi}\ \emph {et~al.}(2010)\citenamefont {Ravi},
  \citenamefont {Lee}, \citenamefont {Sharma}, \citenamefont {Werth},\ and\
  \citenamefont {Rangwala}}]{Ravi2010}%
  \BibitemOpen
  \bibfield  {author} {\bibinfo {author} {\bibfnamefont {K.}~\bibnamefont
  {Ravi}}, \bibinfo {author} {\bibfnamefont {S.}~\bibnamefont {Lee}}, \bibinfo
  {author} {\bibfnamefont {A.}~\bibnamefont {Sharma}}, \bibinfo {author}
  {\bibfnamefont {G.}~\bibnamefont {Werth}}, \ and\ \bibinfo {author}
  {\bibfnamefont {S.~A.}\ \bibnamefont {Rangwala}},\ }\href {\doibase
  arXiv:1009.4515v1} {\bibfield  {journal} {\bibinfo  {journal}
  {arxiv-preprint}\ } (\bibinfo {year} {2010}),\ arXiv:1009.4515v1}\BibitemShut
  {NoStop}%
\bibitem [{\citenamefont {C\^{o}t\'{e}}\ and\ \citenamefont
  {Dalgarno}(2000)}]{Cote2000}%
  \BibitemOpen
  \bibfield  {author} {\bibinfo {author} {\bibfnamefont {R.}~\bibnamefont
  {C\^{o}t\'{e}}}\ and\ \bibinfo {author} {\bibfnamefont {A.}~\bibnamefont
  {Dalgarno}},\ }\href {\doibase 10.1103/PhysRevA.62.012709} {\bibfield
  {journal} {\bibinfo  {journal} {Phys. Rev. A}\ }\textbf {\bibinfo {volume}
  {62}},\ \bibinfo {eid} {012709} (\bibinfo {year} {2000})}\BibitemShut
  {NoStop}%
\bibitem [{\citenamefont {C\^{o}t\'{e}}\ \emph {et~al.}(2002)\citenamefont
  {C\^{o}t\'{e}}, \citenamefont {Kharchenko},\ and\ \citenamefont
  {Lukin}}]{Cote2002}%
  \BibitemOpen
  \bibfield  {author} {\bibinfo {author} {\bibfnamefont {R.}~\bibnamefont
  {C\^{o}t\'{e}}}, \bibinfo {author} {\bibfnamefont {V.}~\bibnamefont
  {Kharchenko}}, \ and\ \bibinfo {author} {\bibfnamefont {M.~D.}\ \bibnamefont
  {Lukin}},\ }\href {\doibase 10.1103/PhysRevLett.89.093001} {\bibfield
  {journal} {\bibinfo  {journal} {Phys. Rev. Lett.}\ }\textbf {\bibinfo
  {volume} {89}},\ \bibinfo {eid} {093001} (\bibinfo {year}
  {2002})}\BibitemShut {NoStop}%
\bibitem [{\citenamefont {Idziaszek}\ \emph {et~al.}(2007)\citenamefont
  {Idziaszek}, \citenamefont {Calarco},\ and\ \citenamefont
  {Zoller}}]{Idziaszek2007}%
  \BibitemOpen
  \bibfield  {author} {\bibinfo {author} {\bibfnamefont {Z.}~\bibnamefont
  {Idziaszek}}, \bibinfo {author} {\bibfnamefont {T.}~\bibnamefont {Calarco}},
  \ and\ \bibinfo {author} {\bibfnamefont {P.}~\bibnamefont {Zoller}},\ }\href
  {\doibase 10.1103/PhysRevA.76.033409} {\bibfield  {journal} {\bibinfo
  {journal} {Phys. Rev. A}\ }\textbf {\bibinfo {volume} {76}},\ \bibinfo {eid}
  {033409} (\bibinfo {year} {2007})}\BibitemShut {NoStop}%
\bibitem [{\citenamefont {Kollath}\ \emph {et~al.}(2007)\citenamefont
  {Kollath}, \citenamefont {K\"{o}hl},\ and\ \citenamefont
  {Giamarchi}}]{Kollath2007}%
  \BibitemOpen
  \bibfield  {author} {\bibinfo {author} {\bibfnamefont {C.}~\bibnamefont
  {Kollath}}, \bibinfo {author} {\bibfnamefont {M.}~\bibnamefont {K\"{o}hl}}, \
  and\ \bibinfo {author} {\bibfnamefont {T.}~\bibnamefont {Giamarchi}},\ }\href
  {\doibase 10.1103/PhysRevA.76.063602} {\bibfield  {journal} {\bibinfo
  {journal} {Phys. Rev. A}\ }\textbf {\bibinfo {volume} {76}},\ \bibinfo {eid}
  {063602} (\bibinfo {year} {2007})}\BibitemShut {NoStop}%
\bibitem [{\citenamefont {Idziaszek}\ \emph {et~al.}(2009)\citenamefont
  {Idziaszek}, \citenamefont {Calarco}, \citenamefont {Julienne},\ and\
  \citenamefont {Simoni}}]{Idziaszek2009}%
  \BibitemOpen
  \bibfield  {author} {\bibinfo {author} {\bibfnamefont {Z.}~\bibnamefont
  {Idziaszek}}, \bibinfo {author} {\bibfnamefont {T.}~\bibnamefont {Calarco}},
  \bibinfo {author} {\bibfnamefont {P.~S.}\ \bibnamefont {Julienne}}, \ and\
  \bibinfo {author} {\bibfnamefont {A.}~\bibnamefont {Simoni}},\ }\href
  {\doibase 10.1103/PhysRevA.79.010702} {\bibfield  {journal} {\bibinfo
  {journal} {Phys. Rev. A}\ }\textbf {\bibinfo {volume} {79}},\ \bibinfo {eid}
  {010702} (\bibinfo {year} {2009})}\BibitemShut {NoStop}%
\bibitem [{\citenamefont {Sherkunov}\ \emph {et~al.}(2009)\citenamefont
  {Sherkunov}, \citenamefont {Muzykantskii}, \citenamefont {d'Ambrumenil},\
  and\ \citenamefont {Simons}}]{Sherkunov2009}%
  \BibitemOpen
  \bibfield  {author} {\bibinfo {author} {\bibfnamefont {Y.}~\bibnamefont
  {Sherkunov}}, \bibinfo {author} {\bibfnamefont {B.}~\bibnamefont
  {Muzykantskii}}, \bibinfo {author} {\bibfnamefont {N.}~\bibnamefont
  {d'Ambrumenil}}, \ and\ \bibinfo {author} {\bibfnamefont {B.~D.}\
  \bibnamefont {Simons}},\ }\href {\doibase 10.1103/PhysRevA.79.023604}
  {\bibfield  {journal} {\bibinfo  {journal} {Phys. Rev. A}\ }\textbf {\bibinfo
  {volume} {79}},\ \bibinfo {eid} {023604} (\bibinfo {year}
  {2009})}\BibitemShut {NoStop}%
\bibitem [{\citenamefont {Doerk}\ \emph {et~al.}(2010)\citenamefont {Doerk},
  \citenamefont {Idziaszek},\ and\ \citenamefont {Calarco}}]{Doerk2010}%
  \BibitemOpen
  \bibfield  {author} {\bibinfo {author} {\bibfnamefont {H.}~\bibnamefont
  {Doerk}}, \bibinfo {author} {\bibfnamefont {Z.}~\bibnamefont {Idziaszek}}, \
  and\ \bibinfo {author} {\bibfnamefont {T.}~\bibnamefont {Calarco}},\ }\href
  {\doibase 10.1103/PhysRevA.81.012708} {\bibfield  {journal} {\bibinfo
  {journal} {Phys. Rev. A}\ }\textbf {\bibinfo {volume} {81}},\ \bibinfo
  {pages} {012708} (\bibinfo {year} {2010})}\BibitemShut {NoStop}%
\bibitem [{\citenamefont {Gao}(2010)}]{Gao2010}%
  \BibitemOpen
  \bibfield  {author} {\bibinfo {author} {\bibfnamefont {B.}~\bibnamefont
  {Gao}},\ }\href {\doibase 10.1103/PhysRevLett.104.213201} {\bibfield
  {journal} {\bibinfo  {journal} {Phys. Rev. Lett.}\ }\textbf {\bibinfo
  {volume} {104}},\ \bibinfo {pages} {213201} (\bibinfo {year}
  {2010})}\BibitemShut {NoStop}%
\bibitem [{\citenamefont {Goold}\ \emph {et~al.}(2010)\citenamefont {Goold},
  \citenamefont {Doerk}, \citenamefont {Idziaszek}, \citenamefont {Calarco},\
  and\ \citenamefont {Busch}}]{Goold2010}%
  \BibitemOpen
  \bibfield  {author} {\bibinfo {author} {\bibfnamefont {J.}~\bibnamefont
  {Goold}}, \bibinfo {author} {\bibfnamefont {H.}~\bibnamefont {Doerk}},
  \bibinfo {author} {\bibfnamefont {Z.}~\bibnamefont {Idziaszek}}, \bibinfo
  {author} {\bibfnamefont {T.}~\bibnamefont {Calarco}}, \ and\ \bibinfo
  {author} {\bibfnamefont {T.}~\bibnamefont {Busch}},\ }\href {\doibase
  10.1103/PhysRevA.81.041601} {\bibfield  {journal} {\bibinfo  {journal} {Phys.
  Rev. A}\ }\textbf {\bibinfo {volume} {81}},\ \bibinfo {pages} {041601}
  (\bibinfo {year} {2010})}\BibitemShut {NoStop}%
\bibitem [{\citenamefont {Idziaszek}\ \emph {et~al.}(2010)\citenamefont
  {Idziaszek}, \citenamefont {Calarco},\ and\ \citenamefont
  {Zoller}}]{Idziaszek2010}%
  \BibitemOpen
  \bibfield  {author} {\bibinfo {author} {\bibfnamefont {Z.}~\bibnamefont
  {Idziaszek}}, \bibinfo {author} {\bibfnamefont {T.}~\bibnamefont {Calarco}},
  \ and\ \bibinfo {author} {\bibfnamefont {P.}~\bibnamefont {Zoller}},\ }\href
  {\doibase arXiv:1008.1858v1} {\bibfield  {journal} {\bibinfo  {journal}
  {arxiv-preprint}\ } (\bibinfo {year} {2010}),\ arXiv:1008.1858v1}\BibitemShut
  {NoStop}%
\bibitem [{\citenamefont {Krych}\ \emph {et~al.}(2010)\citenamefont {Krych},
  \citenamefont {Skomorowski}, \citenamefont {Pawlowski}, \citenamefont
  {Moszynski},\ and\ \citenamefont {Idziaszek}}]{Krych2010}%
  \BibitemOpen
  \bibfield  {author} {\bibinfo {author} {\bibfnamefont {M.}~\bibnamefont
  {Krych}}, \bibinfo {author} {\bibfnamefont {W.}~\bibnamefont {Skomorowski}},
  \bibinfo {author} {\bibfnamefont {F.}~\bibnamefont {Pawlowski}}, \bibinfo
  {author} {\bibfnamefont {R.}~\bibnamefont {Moszynski}}, \ and\ \bibinfo
  {author} {\bibfnamefont {Z.}~\bibnamefont {Idziaszek}},\ }\href {\doibase
  arXiv:1008.0840v1} {\bibfield  {journal} {\bibinfo  {journal}
  {arxiv-preprint}\ } (\bibinfo {year} {2010}),\ arXiv:1008.0840v1}\BibitemShut
  {NoStop}%
\bibitem [{\citenamefont {Makarov}\ \emph {et~al.}(2003)\citenamefont
  {Makarov}, \citenamefont {C\^ot\'e}, \citenamefont {Michels},\ and\
  \citenamefont {Smith}}]{Makarov2003}%
  \BibitemOpen
  \bibfield  {author} {\bibinfo {author} {\bibfnamefont {O.~P.}\ \bibnamefont
  {Makarov}}, \bibinfo {author} {\bibfnamefont {R.}~\bibnamefont {C\^ot\'e}},
  \bibinfo {author} {\bibfnamefont {H.}~\bibnamefont {Michels}}, \ and\
  \bibinfo {author} {\bibfnamefont {W.~W.}\ \bibnamefont {Smith}},\ }\href
  {\doibase 10.1103/PhysRevA.67.042705} {\bibfield  {journal} {\bibinfo
  {journal} {Phys. Rev. A}\ }\textbf {\bibinfo {volume} {67}},\ \bibinfo
  {pages} {042705} (\bibinfo {year} {2003})}\BibitemShut {NoStop}%
\bibitem [{\citenamefont {Hudson}(2009)}]{Hudson2009}%
  \BibitemOpen
  \bibfield  {author} {\bibinfo {author} {\bibfnamefont {E.~R.}\ \bibnamefont
  {Hudson}},\ }\href {\doibase 10.1103/PhysRevA.79.032716} {\bibfield
  {journal} {\bibinfo  {journal} {Phys. Rev. A}\ }\textbf {\bibinfo {volume}
  {79}},\ \bibinfo {pages} {032716} (\bibinfo {year} {2009})}\BibitemShut
  {NoStop}%
\bibitem [{\citenamefont {Langevin}(1905)}]{Langevin1905}%
  \BibitemOpen
  \bibfield  {author} {\bibinfo {author} {\bibfnamefont {P.}~\bibnamefont
  {Langevin}},\ }\href@noop {} {\bibfield  {journal} {\bibinfo  {journal} {Ann.
  Chim. Phys.}\ }\textbf {\bibinfo {volume} {5}},\ \bibinfo {pages} {245}
  (\bibinfo {year} {1905})}\BibitemShut {NoStop}%
\bibitem [{\citenamefont {Albritton}\ \emph {et~al.}(1968)\citenamefont
  {Albritton}, \citenamefont {Miller}, \citenamefont {Martin},\ and\
  \citenamefont {McDaniel}}]{Albritton1968}%
  \BibitemOpen
  \bibfield  {author} {\bibinfo {author} {\bibfnamefont {D.~L.}\ \bibnamefont
  {Albritton}}, \bibinfo {author} {\bibfnamefont {T.~M.}\ \bibnamefont
  {Miller}}, \bibinfo {author} {\bibfnamefont {D.~W.}\ \bibnamefont {Martin}},
  \ and\ \bibinfo {author} {\bibfnamefont {E.~W.}\ \bibnamefont {McDaniel}},\
  }\href {\doibase 10.1103/PhysRev.171.94} {\bibfield  {journal} {\bibinfo
  {journal} {Phys. Rev.}\ }\textbf {\bibinfo {volume} {171}},\ \bibinfo {pages}
  {94} (\bibinfo {year} {1968})}\BibitemShut {NoStop}%
\bibitem [{\citenamefont {Major}\ and\ \citenamefont
  {Dehmelt}(1968)}]{Major1968}%
  \BibitemOpen
  \bibfield  {author} {\bibinfo {author} {\bibfnamefont {F.~G.}\ \bibnamefont
  {Major}}\ and\ \bibinfo {author} {\bibfnamefont {H.~G.}\ \bibnamefont
  {Dehmelt}},\ }\href {\doibase 10.1103/PhysRev.170.91} {\bibfield  {journal}
  {\bibinfo  {journal} {Phys. Rev.}\ }\textbf {\bibinfo {volume} {170}},\
  \bibinfo {pages} {91} (\bibinfo {year} {1968})}\BibitemShut {NoStop}%
\bibitem [{\citenamefont {Ridinger}\ and\ \citenamefont
  {Davidson}(2007)}]{Ridinger2007}%
  \BibitemOpen
  \bibfield  {author} {\bibinfo {author} {\bibfnamefont {A.}~\bibnamefont
  {Ridinger}}\ and\ \bibinfo {author} {\bibfnamefont {N.}~\bibnamefont
  {Davidson}},\ }\href {\doibase 10.1103/PhysRevA.76.013421} {\bibfield
  {journal} {\bibinfo  {journal} {Phys. Rev. A}\ }\textbf {\bibinfo {volume}
  {76}},\ \bibinfo {pages} {013421} (\bibinfo {year} {2007})}\BibitemShut
  {NoStop}%
\bibitem [{\citenamefont {Ridinger}\ and\ \citenamefont
  {Weiss}(2009)}]{Ridinger2009}%
  \BibitemOpen
  \bibfield  {author} {\bibinfo {author} {\bibfnamefont {A.}~\bibnamefont
  {Ridinger}}\ and\ \bibinfo {author} {\bibfnamefont {C.}~\bibnamefont
  {Weiss}},\ }\href {\doibase 10.1103/PhysRevA.79.013414} {\bibfield  {journal}
  {\bibinfo  {journal} {Phys. Rev. A}\ }\textbf {\bibinfo {volume} {79}},\
  \bibinfo {pages} {013414} (\bibinfo {year} {2009})}\BibitemShut {NoStop}%
\bibitem [{\citenamefont {Moriwaki}\ and\ \citenamefont
  {Shimizu}(1998)}]{Moriwaki1998}%
  \BibitemOpen
  \bibfield  {author} {\bibinfo {author} {\bibfnamefont {Y.}~\bibnamefont
  {Moriwaki}}\ and\ \bibinfo {author} {\bibfnamefont {T.}~\bibnamefont
  {Shimizu}},\ }\href {\doibase 10.1143/JJAP.37.344} {\bibfield  {journal}
  {\bibinfo  {journal} {Japanese Journal of Applied Physics}\ }\textbf
  {\bibinfo {volume} {37}},\ \bibinfo {pages} {344} (\bibinfo {year}
  {1998})}\BibitemShut {NoStop}%
\bibitem [{\citenamefont {Green}\ \emph {et~al.}(2007)\citenamefont {Green},
  \citenamefont {Wodin}, \citenamefont {DeVoe}, \citenamefont {Fierlinger},
  \citenamefont {Flatt}, \citenamefont {Gratta}, \citenamefont {LePort},
  \citenamefont {Montero~D\'\i{}ez}, \citenamefont {Neilson}, \citenamefont
  {O'Sullivan}, \citenamefont {Pocar}, \citenamefont {Waldman}, \citenamefont
  {Leonard}, \citenamefont {Piepke}, \citenamefont {Hargrove}, \citenamefont
  {Sinclair}, \citenamefont {Strickland}, \citenamefont {Fairbank},
  \citenamefont {Hall}, \citenamefont {Mong}, \citenamefont {Moe},
  \citenamefont {Farine}, \citenamefont {Hallman}, \citenamefont {Virtue},
  \citenamefont {Baussan}, \citenamefont {Martin}, \citenamefont {Schenker},
  \citenamefont {Vuilleumier}, \citenamefont {Vuilleumier}, \citenamefont
  {Weber}, \citenamefont {Breidenbach}, \citenamefont {Conley}, \citenamefont
  {Hall}, \citenamefont {Hodgson}, \citenamefont {Mackay}, \citenamefont
  {Odian}, \citenamefont {Prescott}, \citenamefont {Rowson}, \citenamefont
  {Skarpaas},\ and\ \citenamefont {Wamba}}]{Green2007}%
  \BibitemOpen
  \bibfield  {author} {\bibinfo {author} {\bibfnamefont {M.}~\bibnamefont
  {Green}}, \bibinfo {author} {\bibfnamefont {J.}~\bibnamefont {Wodin}},
  \bibinfo {author} {\bibfnamefont {R.}~\bibnamefont {DeVoe}}, \bibinfo
  {author} {\bibfnamefont {P.}~\bibnamefont {Fierlinger}}, \bibinfo {author}
  {\bibfnamefont {B.}~\bibnamefont {Flatt}}, \bibinfo {author} {\bibfnamefont
  {G.}~\bibnamefont {Gratta}}, \bibinfo {author} {\bibfnamefont
  {F.}~\bibnamefont {LePort}}, \bibinfo {author} {\bibfnamefont
  {M.}~\bibnamefont {Montero~D\'\i{}ez}}, \bibinfo {author} {\bibfnamefont
  {R.}~\bibnamefont {Neilson}}, \bibinfo {author} {\bibfnamefont
  {K.}~\bibnamefont {O'Sullivan}}, \bibinfo {author} {\bibfnamefont
  {A.}~\bibnamefont {Pocar}}, \bibinfo {author} {\bibfnamefont
  {S.}~\bibnamefont {Waldman}}, \bibinfo {author} {\bibfnamefont {D.~S.}\
  \bibnamefont {Leonard}}, \bibinfo {author} {\bibfnamefont {A.}~\bibnamefont
  {Piepke}}, \bibinfo {author} {\bibfnamefont {C.}~\bibnamefont {Hargrove}},
  \bibinfo {author} {\bibfnamefont {D.}~\bibnamefont {Sinclair}}, \bibinfo
  {author} {\bibfnamefont {V.}~\bibnamefont {Strickland}}, \bibinfo {author}
  {\bibfnamefont {W.}~\bibnamefont {Fairbank}}, \bibinfo {author}
  {\bibfnamefont {K.}~\bibnamefont {Hall}}, \bibinfo {author} {\bibfnamefont
  {B.}~\bibnamefont {Mong}}, \bibinfo {author} {\bibfnamefont {M.}~\bibnamefont
  {Moe}}, \bibinfo {author} {\bibfnamefont {J.}~\bibnamefont {Farine}},
  \bibinfo {author} {\bibfnamefont {D.}~\bibnamefont {Hallman}}, \bibinfo
  {author} {\bibfnamefont {C.}~\bibnamefont {Virtue}}, \bibinfo {author}
  {\bibfnamefont {E.}~\bibnamefont {Baussan}}, \bibinfo {author} {\bibfnamefont
  {Y.}~\bibnamefont {Martin}}, \bibinfo {author} {\bibfnamefont
  {D.}~\bibnamefont {Schenker}}, \bibinfo {author} {\bibfnamefont {J.-L.}\
  \bibnamefont {Vuilleumier}}, \bibinfo {author} {\bibfnamefont {J.-M.}\
  \bibnamefont {Vuilleumier}}, \bibinfo {author} {\bibfnamefont
  {P.}~\bibnamefont {Weber}}, \bibinfo {author} {\bibfnamefont
  {M.}~\bibnamefont {Breidenbach}}, \bibinfo {author} {\bibfnamefont
  {R.}~\bibnamefont {Conley}}, \bibinfo {author} {\bibfnamefont
  {C.}~\bibnamefont {Hall}}, \bibinfo {author} {\bibfnamefont {J.}~\bibnamefont
  {Hodgson}}, \bibinfo {author} {\bibfnamefont {D.}~\bibnamefont {Mackay}},
  \bibinfo {author} {\bibfnamefont {A.}~\bibnamefont {Odian}}, \bibinfo
  {author} {\bibfnamefont {C.~Y.}\ \bibnamefont {Prescott}}, \bibinfo {author}
  {\bibfnamefont {P.~C.}\ \bibnamefont {Rowson}}, \bibinfo {author}
  {\bibfnamefont {K.}~\bibnamefont {Skarpaas}}, \ and\ \bibinfo {author}
  {\bibfnamefont {K.}~\bibnamefont {Wamba}},\ }\href {\doibase
  10.1103/PhysRevA.76.023404} {\bibfield  {journal} {\bibinfo  {journal} {Phys.
  Rev. A}\ }\textbf {\bibinfo {volume} {76}},\ \bibinfo {pages} {023404}
  (\bibinfo {year} {2007})}\BibitemShut {NoStop}%
\bibitem [{\citenamefont {DeVoe}(2009)}]{Devoe2009}%
  \BibitemOpen
  \bibfield  {author} {\bibinfo {author} {\bibfnamefont {R.~G.}\ \bibnamefont
  {DeVoe}},\ }\href {\doibase 10.1103/PhysRevLett.102.063001} {\bibfield
  {journal} {\bibinfo  {journal} {Phys. Rev. Lett.}\ }\textbf {\bibinfo
  {volume} {102}},\ \bibinfo {eid} {063001} (\bibinfo {year}
  {2009})}\BibitemShut {NoStop}%
\bibitem [{\citenamefont {Kim}(1997)}]{Kim1997}%
  \BibitemOpen
  \bibfield  {author} {\bibinfo {author} {\bibfnamefont {T.}~\bibnamefont
  {Kim}},\ }\emph {\bibinfo {title} {Buffer gas cooling of ions in a radio
  frequency quadrupole ion guide}},\ \href@noop {} {Ph.D. thesis},\ \bibinfo
  {school} {McGill University} (\bibinfo {year} {1997})\BibitemShut {NoStop}%
\bibitem [{\citenamefont {Kellerbauer}\ \emph {et~al.}(2001)\citenamefont
  {Kellerbauer}, \citenamefont {Kim}, \citenamefont {Moore},\ and\
  \citenamefont {Varfa}}]{Kellerbauer2001}%
  \BibitemOpen
  \bibfield  {author} {\bibinfo {author} {\bibfnamefont {A.}~\bibnamefont
  {Kellerbauer}}, \bibinfo {author} {\bibfnamefont {T.}~\bibnamefont {Kim}},
  \bibinfo {author} {\bibfnamefont {R.}~\bibnamefont {Moore}}, \ and\ \bibinfo
  {author} {\bibfnamefont {P.}~\bibnamefont {Varfa}},\ }\href@noop {}
  {\bibfield  {journal} {\bibinfo  {journal} {Nucl. Instrum. Methods Phys.
  Res., Sect. A}\ }\textbf {\bibinfo {volume} {469}},\ \bibinfo {pages} {276 }
  (\bibinfo {year} {2001})}\BibitemShut {NoStop}%
\bibitem [{\citenamefont {Schwarz}(2006)}]{Schwarz2006}%
  \BibitemOpen
  \bibfield  {author} {\bibinfo {author} {\bibfnamefont {S.}~\bibnamefont
  {Schwarz}},\ }\href@noop {} {\bibfield  {journal} {\bibinfo  {journal} {Nucl.
  Instrum. Methods Phys. Res., Sect. A}\ }\textbf {\bibinfo {volume} {566}},\
  \bibinfo {pages} {233} (\bibinfo {year} {2006})}\BibitemShut {NoStop}%
\bibitem [{\citenamefont {Wineland}\ \emph {et~al.}(1998)\citenamefont
  {Wineland}, \citenamefont {Monroe}, \citenamefont {Itano}, \citenamefont
  {Leibfried}, \citenamefont {King},\ and\ \citenamefont
  {Meekhof}}]{Wineland1998}%
  \BibitemOpen
  \bibfield  {author} {\bibinfo {author} {\bibfnamefont {D.}~\bibnamefont
  {Wineland}}, \bibinfo {author} {\bibfnamefont {C.}~\bibnamefont {Monroe}},
  \bibinfo {author} {\bibfnamefont {W.~M.}\ \bibnamefont {Itano}}, \bibinfo
  {author} {\bibfnamefont {D.}~\bibnamefont {Leibfried}}, \bibinfo {author}
  {\bibfnamefont {B.~E.}\ \bibnamefont {King}}, \ and\ \bibinfo {author}
  {\bibfnamefont {D.~M.}\ \bibnamefont {Meekhof}},\ }\href@noop {} {\bibfield
  {journal} {\bibinfo  {journal} {J.Res.Natl.Inst.Stand.Tech.}\ }\textbf
  {\bibinfo {volume} {103}},\ \bibinfo {pages} {259} (\bibinfo {year}
  {1998})}\BibitemShut {NoStop}%
\bibitem [{\citenamefont {Leibfried}\ \emph {et~al.}(2003)\citenamefont
  {Leibfried}, \citenamefont {Blatt}, \citenamefont {Monroe},\ and\
  \citenamefont {Wineland}}]{Leibfried2003}%
  \BibitemOpen
  \bibfield  {author} {\bibinfo {author} {\bibfnamefont {D.}~\bibnamefont
  {Leibfried}}, \bibinfo {author} {\bibfnamefont {R.}~\bibnamefont {Blatt}},
  \bibinfo {author} {\bibfnamefont {C.}~\bibnamefont {Monroe}}, \ and\ \bibinfo
  {author} {\bibfnamefont {D.}~\bibnamefont {Wineland}},\ }\href {\doibase
  10.1103/RevModPhys.75.281} {\bibfield  {journal} {\bibinfo  {journal} {Rev.
  Mod. Phys.}\ }\textbf {\bibinfo {volume} {75}},\ \bibinfo {pages} {281}
  (\bibinfo {year} {2003})}\BibitemShut {NoStop}%
\bibitem [{Note1()}]{Note1}%
  \BibitemOpen
  \bibinfo {note} {This assumption is valid for collision energies above $\hbar
  \protect \tmspace +\thinmuskip {.1667em}\Omega _T$.}\BibitemShut {Stop}%
\bibitem [{Note2()}]{Note2}%
  \BibitemOpen
  \bibinfo {note} {Not included in this expression are possible additional
  micromotion terms that are $\pi /2$ out of phase, therefore proportional to
  $\protect \qopname \relax o{sin}(\Omega _T\protect \tmspace +\thinmuskip
  {.1667em}t)$, which can arise, for example, from RF-phase mismatches on
  opposing electrodes. They are added to $\protect \mathaccentV
  {vec}17E{v}_{mm}$ for simulations where specific experimental data is to be
  represented~\cite {Zipkes2010b}.}\BibitemShut {Stop}%
\bibitem [{\citenamefont {Massey}\ and\ \citenamefont
  {Mohr}(1934)}]{Massey1934}%
  \BibitemOpen
  \bibfield  {author} {\bibinfo {author} {\bibfnamefont {H.~S.~W.}\
  \bibnamefont {Massey}}\ and\ \bibinfo {author} {\bibfnamefont {C.~B.~O.}\
  \bibnamefont {Mohr}},\ }\href {\doibase 10.1098/rspa.1934.0042} {\bibfield
  {journal} {\bibinfo  {journal} {Proc. Roy. Soc. A}\ }\textbf {\bibinfo
  {volume} {144}},\ \bibinfo {pages} {188} (\bibinfo {year}
  {1934})}\BibitemShut {NoStop}%
\bibitem [{\citenamefont {C\^ot\'e}(2000)}]{Cote2000b}%
  \BibitemOpen
  \bibfield  {author} {\bibinfo {author} {\bibfnamefont {R.}~\bibnamefont
  {C\^ot\'e}},\ }\href {\doibase 10.1103/PhysRevLett.85.5316} {\bibfield
  {journal} {\bibinfo  {journal} {Phys. Rev. Lett.}\ }\textbf {\bibinfo
  {volume} {85}},\ \bibinfo {pages} {5316} (\bibinfo {year}
  {2000})}\BibitemShut {NoStop}%
\bibitem [{\citenamefont {Zhang}\ \emph {et~al.}(2009)\citenamefont {Zhang},
  \citenamefont {Dalgarno},\ and\ \citenamefont {C\^ot\'e}}]{Zhang2009}%
  \BibitemOpen
  \bibfield  {author} {\bibinfo {author} {\bibfnamefont {P.}~\bibnamefont
  {Zhang}}, \bibinfo {author} {\bibfnamefont {A.}~\bibnamefont {Dalgarno}}, \
  and\ \bibinfo {author} {\bibfnamefont {R.}~\bibnamefont {C\^ot\'e}},\ }\href
  {\doibase 10.1103/PhysRevA.80.030703} {\bibfield  {journal} {\bibinfo
  {journal} {Phys. Rev. A}\ }\textbf {\bibinfo {volume} {80}},\ \bibinfo
  {pages} {030703} (\bibinfo {year} {2009})}\BibitemShut {NoStop}%
\bibitem [{\citenamefont {Vogt}\ and\ \citenamefont
  {Wannier}(1954)}]{Vogt1954}%
  \BibitemOpen
  \bibfield  {author} {\bibinfo {author} {\bibfnamefont {E.}~\bibnamefont
  {Vogt}}\ and\ \bibinfo {author} {\bibfnamefont {G.~H.}\ \bibnamefont
  {Wannier}},\ }\href {\doibase 10.1103/PhysRev.95.1190} {\bibfield  {journal}
  {\bibinfo  {journal} {Phys. Rev.}\ }\textbf {\bibinfo {volume} {95}},\
  \bibinfo {pages} {1190} (\bibinfo {year} {1954})}\BibitemShut {NoStop}%
\bibitem [{\citenamefont {Bl\"umel}\ \emph {et~al.}(1989)\citenamefont
  {Bl\"umel}, \citenamefont {Kappler}, \citenamefont {Quint},\ and\
  \citenamefont {Walther}}]{Bluemel1989}%
  \BibitemOpen
  \bibfield  {author} {\bibinfo {author} {\bibfnamefont {R.}~\bibnamefont
  {Bl\"umel}}, \bibinfo {author} {\bibfnamefont {C.}~\bibnamefont {Kappler}},
  \bibinfo {author} {\bibfnamefont {W.}~\bibnamefont {Quint}}, \ and\ \bibinfo
  {author} {\bibfnamefont {H.}~\bibnamefont {Walther}},\ }\href {\doibase
  10.1103/PhysRevA.40.808} {\bibfield  {journal} {\bibinfo  {journal} {Phys.
  Rev. A}\ }\textbf {\bibinfo {volume} {40}},\ \bibinfo {pages} {808} (\bibinfo
  {year} {1989})}\BibitemShut {NoStop}%
\bibitem [{\citenamefont {Berkeland}\ \emph {et~al.}(1998)\citenamefont
  {Berkeland}, \citenamefont {Miller}, \citenamefont {Bergquist}, \citenamefont
  {Itano},\ and\ \citenamefont {Wineland}}]{Berkeland1998}%
  \BibitemOpen
  \bibfield  {author} {\bibinfo {author} {\bibfnamefont {D.~J.}\ \bibnamefont
  {Berkeland}}, \bibinfo {author} {\bibfnamefont {J.~D.}\ \bibnamefont
  {Miller}}, \bibinfo {author} {\bibfnamefont {J.~C.}\ \bibnamefont
  {Bergquist}}, \bibinfo {author} {\bibfnamefont {W.~M.}\ \bibnamefont
  {Itano}}, \ and\ \bibinfo {author} {\bibfnamefont {D.~J.}\ \bibnamefont
  {Wineland}},\ }\href {\doibase 10.1063/1.367318} {\bibfield  {journal}
  {\bibinfo  {journal} {Journal of Applied Physics}\ }\textbf {\bibinfo
  {volume} {83}},\ \bibinfo {pages} {5025} (\bibinfo {year}
  {1998})}\BibitemShut {NoStop}%
\bibitem [{\citenamefont {Sakurai}(1994)}]{Sakurai1994}%
  \BibitemOpen
  \bibfield  {author} {\bibinfo {author} {\bibfnamefont {J.}~\bibnamefont
  {Sakurai}},\ }\href@noop {} {\emph {\bibinfo {title} {Modern Quantum
  Mechanics}}}\ (\bibinfo  {publisher} {Addison-Wesley},\ \bibinfo {year}
  {1994})\BibitemShut {NoStop}%
\bibitem [{Note3()}]{Note3}%
  \BibitemOpen
  \bibinfo {note} {As most of the inelastic processes require a close encounter
  collision ($l<l_0$), their cross-sections are expected to scale with
  $E^{-1/2}$.}\BibitemShut {Stop}%
\end{thebibliography}
\end{document}